\documentclass{aa}
\usepackage{natbib}
\usepackage[varg]{txfonts}
\usepackage{graphicx}
\usepackage{wasysym}
\usepackage{color}
\usepackage[draft]{hyperref}
\usepackage{url}
\bibpunct{(}{)}{;}{a}{}{,}

\begin{document}

\title{On the varied origins of up-bending breaks in galaxy disks}

\author{Aaron E. Watkins\inst{1}
  \and Jarkko Laine\inst{2}
  \and S\'{e}bastien Comer\'{o}n\inst{1}
  \and Joachim Janz\inst{1,3}
  \and Heikki Salo\inst{1}}

\institute{Astronomy Research Unit, University of Oulu,
  FIN-90014, Finland
 \and Hamburg Sternwarte, Universität Hamburg, 21029,
 Germany,
 \and Finnish Centre of Astronomy with ESO (FINCA), University of
 Turku, V\"{a}is\"{a}l\"{a}ntie 20, 21500 Piikki\"{o}, Finland}

\abstract {}
          {Using a sample of 175 low-inclination galaxies from the
            S$^{4}$G, we investigate the origins of up-bending (Type
            III) breaks in the 3.6 $\mu$m surface brightness
            profiles of disk galaxies.}
          {We re-analyze a sample of previously identified Type~III
            disk break-hosting galaxies using a new, unbiased
            break-finding algorithm, which uncovered many new,
              sometimes subtle disk breaks across the whole sample.
            We classify each break by its likely origin through close
            examination of the galaxy images across wavelengths, and
            compare samples of galaxies separated by their outermost
            identified break types in terms of their stellar
            populations and local environments.}
          {We find that more than half of the confirmed Type~III
            breaks in our sample can be attributed to morphological
            asymmetry in the host galaxies. As these breaks are
              mostly an artifact of the azimuthal averaging process,
              their status as physical ``breaks'' is questionable.
            Such galaxies occupy some of the highest density
            environments in our sample, implying that much of this
            asymmetry is the result of tidal disturbance.
            Additionally, we find that Type~III breaks related to
            extended spiral arms or star formation often host 
              down-bending (Type~II) breaks at larger radius which
              were previously unidentified.  Such galaxies reside in
            the lowest density environments in our sample, in
              line with previous studies that found a lack of Type~II
              breaks in clusters.  Galaxies occupying the highest
            density environments most often show Type~III breaks
            associated with outer spheroidal components.}
          {We find that Type~III breaks in the outer disks of galaxies
            arise most often through environmental influence: either
            tidal disturbance (resulting in disk asymmetry) or heating
            through, e.g., galaxy harrassment (leading to spheroidal
            components).  Galaxies hosting the latter break types also
            show bimodal distributions in central $g-r$ color and
            morphological type, with more than half of such galaxies
            classified as Sa or earlier; this suggests these galaxies
            may be evolving into early-type galaxies.  By
            contrast, we find that Type~III breaks related to
            apparently secular features (e.g., spiral arms) may not
            truly define their hosts' outer disks, as often in such
            galaxies additional significant breaks can be found at
            larger radius.  Given this variety in Type~III break
            origins, we recommend in future break studies making
              a more detailed distinction between break subtypes when
            seeking out, for example, correlations between disk breaks
            and environment, to avoid mixing unlike physical
            phenomena.}

\maketitle

\newpage

\section{Introduction}

The outer regions of disk galaxies are important tracers of galaxy
formation and evolution.  Their radial extent and coherent rotation
make them particularly sensitive to environmental influences, the
effects of which are preserved over long periods by extended
dynamical timescales.  Outer disks also serve as excellent
laboratories for the evolution of galaxies at low mass surface
density, probing the connection between disk stability and star
formation \citep[e.g.,][]{kennicutt89, martin01, thilker07, goddard10}
and the preponderance of stellar migration across disks through
interactions with bars, spiral arms, or satellites
  \citep[e.g.,][]{sellwood02, debattista06, roskar08, minchev12,
    ruizlara17}.  This low mass surface density also yields highly
inefficient star formation \citep{bigiel10}, despite gas dominating
the mass budget \citep[e.g.][]{broeils97, sancisi08}, a situation
reflective of many dwarf galaxies \citep[e.g.,][]{vanzee97, bigiel10,
  elmegreen15}.

Stellar disks often show abrupt truncations or anti-truncations
in their radial surface brightness profiles at extended radii
\citep[e.g.,][]{freeman70, vanderkruit87, pohlen06, erwin08,
  gutierrez11, martinnavarro12, laine14}, giving clues to how mass
is distributed across the disk throughout the galaxy's lifetime.
Since the seminal work by \citet{pohlen06} and \citet{erwin08}, disk
galaxies have been separated into three major classes \citep[following
  and expanding upon the system devised by][]{freeman70}: Type~I
disks, with surface brightness profiles described well by a single
exponential decline; Type~II disks, best described by a broken
exponential with a steeper decline in the disk outskirts; and Type~III
disks, described by a broken exponential with a shallower decline in
the outskirts.  Type~II disk breaks, the most common \citep{pohlen06,
  erwin08, laine14}, are also perhaps the most well-understood; often
their presence is attributed to a combination of a dynamical star
formation threshold \citep[e.g.,][]{kennicutt89, martin01}, which
truncates the young stellar disk, and radial migration, which
populates the disk beyond the break radius with evolved stars
\citep[e.g.,][]{sellwood02, debattista06, bakos08, roskar08,
  minchev11, munozmateos13}.

The most poorly understood breaks are those of Type~III, for which a
wide variety of explanations have been proffered.  \citet{younger07},
for example, showed that Type~III breaks can arise through accretion
of a gas-rich companion, while \citet{laurikainen01} found that
shallow outer profiles are often introduced into Messier 51-like
interacting pairs through long-lived redistributions of mass.  Kinematic
heating of stellar populations through major mergers
\citep{borlaff14}, bombardment by cold dark matter substructure
\citep{kazantzidis09}, and galaxy harassment \citep{roediger12} may
form some Type~III disk breaks as well.  Tidally induced isophotal
asymmetry may also often lead to Type~III disk breaks \citep{erwin05,
  laine14}, and because S0 galaxies host such breaks more frequently
than late-type spirals, regardless of environment
\citep[e.g.][]{erwin05, gutierrez11, Ilyina12, maltby15}, Type~III
disk break formation and S0 formation may be linked.  In some cases
\citep[denoted Type~IIIh or IIIs;][]{pohlen06, erwin08}, Type~III
breaks may instead reflect a transition to a kinematically hot stellar
component, such as the stellar halo \citep{martinnavarro12,
  martinnavarro14, peters17} or thick disk \citep{comeron12,
  comeron18}.

In the realm of star formation, Laine et al. (2016; hereafter, L16)
found that disk scalelengths beyond Type~III breaks in their sample
were longest in the \emph{u}-band, suggesting predominantly young
trans-break populations.  Similarly, \citet{wang18} showed that Type
III disks are more prevalent in low-spin, HI-rich galaxies, suggesting
that star formation induced through gas accretion results in Type~III
breaks.  Enhanced star formation in the \emph{inner} disk may also
yield steeper inner disk profiles than outer disk profiles, thereby
producing anti-truncations \citep{hunter06}.  Some Type~III breaks
have been linked to outer rings or pseudorings \citep{erwin05,
  pohlen06}, and in simulations have been linked to scattering by bars
\citep{herpich17}, implying that internal secular processes can also
lead to Type~III breaks.  In this vein, Type~III breaks may also
  arise through a combination of accretion events and outward radial
  migration \citep[e.g.,][]{ruizlara17}, or migration induced by
  infall into a cluster environment \citep[e.g.,][]{clarke17}.

It is worth summarizing these possibilities: Type~III breaks may
result from minor mergers, major mergers, flyby interactions, dark
matter substructure, harassment, stellar halos, thick disks, enhanced
outer disk star formation, enhanced inner disk star formation, outer
rings, bar scattering, radial migration, and other internal
  secular processes in both isolated and interacting galaxies.  These
explanations span nearly every possible aspect of disk galaxy
evolution.

Comparitive studies of disk breaks, by necessity, do not take this
diversity into account, assessing only the relative frequencies or
average properties of Type~I, Type~II, and Type~III disk break hosts
\citep[e.g.,][]{pohlen06, bakos08, gutierrez11, roediger12, laine14,
  maltby15, wang18}.  If each type of break can be created through
such a diverse array of mechanisms, however, this approach may lack
the complexity necessary for a complete understanding of disk breaks.
Indeed, this diversity may be one reason different disk break studies
come to conflicting conclusions: for example, one might expect that if
most Type~III disk breaks resulted from enhanced outer disk star
formation (L16, Wang et al. 2018), evidence of this would appear
through frequent blueward color gradients in Type~III disks.  Yet
\citet{bakos08} and \citet{zheng15} found that, on average, Type~III
color profiles are flat.  Also, if many Type~III disk breaks form
through tidal disturbance \citep[e.g.,][]{erwin05, laine14}, their
hosts should be more often found in dense environments.  Yet several
studies of late-type galaxies either failed to find such a correlation
\citep[e.g.,][]{pohlen06, maltby12}, or found only a weak such
correlation \citep[e.g.,][]{laine14}.

Therefore, to elucidate the formation mechanisms of Type~III disk
breaks, we perform a detailed re-analysis of the 3.6 $\mu$m
  images of Type~III break-hosting galaxies in the \emph{Spitzer}
Survey of Stellar Structure in Galaxies \citep[S$^{4}$G;][]{sheth10},
as previously classified by L16.  We introduce a new unbiased
break-finding algorithm and incorporate the detailed morphological
classifications of \citet{buta15} into both the break identification
and interpretation.  We then classify each identified break using a
new system of break subclassifications tailored to the apparent
physical origin of each break, and explore the stellar populations of
the break hosts and correlations between the break subcategories and
the local environment.  In Section 2, we discuss the datasets used in
this study, as well as the chosen sample of galaxies.  In Section 3,
we discuss our break-finding methodology, and present an overview of
our break classification scheme.  Section 4 presents the results of
our study, including color profiles by break type and environmental
correlation with break type.  Section 5 provides a discussion of the
potential physical origins and data-driven causes of these break
subcategories, and Section 6 provides a summary.

\section{Sample and Data}

To facilitate comparisons with previous work, we selected our sample
from that defined by L16, an expansion of the sample defined by
  \citet{laine14} to include later Hubble types and lower stellar
  masses.  In summary, the sample was chosen from the the S$^{4}$G
\citep{sheth10} and the Near InfraRed S0-Sa Survey
\citep[NIRS0S;][]{laurikainen11}, including only those galaxies with
Hubble types $-2 \leq T \leq 9$ \citep[using the near-IR
  classifications from][]{laurikainen11, buta15} and minor/major axis
ratios $b/a > 0.5$ (as measured from the outer disk).  This sample
covers Local Universe galaxies with a wide range of stellar masses ($8
\lesssim \log_{10}(M_{*}/M_{\odot}) \lesssim 11$, albeit with
  systematically lower masses for Hubble types T$\gtrsim 5$; see
  Fig.~1 of L16), and through the axis ratio restriction includes
only those galaxies for which the features in the surface brightness
profiles can be tied to morphological features in the disks.

For the present study, we re-examined the 175 (of 753 total)
galaxies L16 classified as having Type~III breaks in one or more of
the following bands: 3.6$\mu$m from the S$^{4}$G survey;
\emph{K}$_{s}$-band from the NIRS0S survey; \emph{u}, \emph{g},
\emph{r}, \emph{i}, and \emph{z}-band imaging from a re-reduction of
Sloan Digital Sky Survey \citep[SDSS, specifically DR7 \&
  DR8;][]{abazajian09, aihara11} images done by \citet{knapen14}; and
\emph{g}-band imaging from the Liverpool Telescope for 111 S$^{4}$G
galaxies not found in SDSS \citep{knapen14}.  The Type~III sub-sample
contains no NIRS0S survey galaxies, as many NIRS0S galaxies were
  removed from the original sample of \citet{laine14} by L16 through a
  combination of a tighter restriction on image depth and a revised
  T-type selection (T$\geq$-2, compared to T$\geq$-3).  Additionally,
we included any ambiguous Type~III break classifications as well,
  for the sake of completeness and a larger sample size.  For
  example, 5 galaxies in the present sample were classified as
Type~III disks only in optical bands, and 39 were given hybrid
classifications (e.g., Type~III+II).  Color information from SDSS
  imaging is available for only 113 galaxies in our sample.

As L16, we use galaxy centers, position angles, axial ratios,
background sky levels, and background standard deviations from
\citet{salo15}.  We utilize the near-IR morphological classifications
from \citet{buta15}, as well as the foreground star, background
galaxy, and artifact masks from \citet{munozmateos15}, enlarged
  through dilation with a 3 pixel radius top-hat kernel in order to
  avoid spurious breaks at low surface brightness (see the
  Appendix).

While we used only the 3.6 $\mu$m images in our break-finding
  analysis, we measured surface brightness profiles in optical bands
  as well to derive radial color profiles.  In all bands, we measured
  surface brightness profiles using elliptical apertures with fixed
centers, position angles, and axial ratios derived from the 3.6 $\mu$m
images using the IRAF\footnote{IRAF is distributed by the National
  Optical Astronomy Observatory, which is operated by the Association
  of Universities for Research in Astronomy (AURA), Inc., under
  cooperative agreement with the National Science Foundation.} ellipse
package \citep{jedrzejewski87}, with orientations and axial ratios
reflective of each galaxy's outer disk.  Specifically, we measured the
mean flux within each annular aperture $i$, $\bar{f}_{i}$ (with fixed
widths of 1.5\arcsec, interpolating across masked regions), and
derived the surface brightness as
\begin{equation}\label{eq:sb}
  \mu_{i} = -2.5\log_{10}(\bar{f}_{i}) + \textrm{ZP} +
  2.5\log_{10}(p^{2})
\end{equation}
where ZP is the surface brightness zero-point for the given band and
$p$ is the pixel scale in arcsec px$^{-1}$.

We describe our break-finding algorithm using these one-dimensional
profiles in the following section.  We also provide a detailed
assessment of methodological biases on the derivation of the surface
brightness profiles in the Appendix, including the effects of masking,
interpolation of flux across masks, and the use of the mean flux
vs. the median flux in each aperture.

\section{Break identification and classification}

\subsection{Break-finding algorithm}

\begin{figure*}
  \centering
  \includegraphics[scale=0.95]{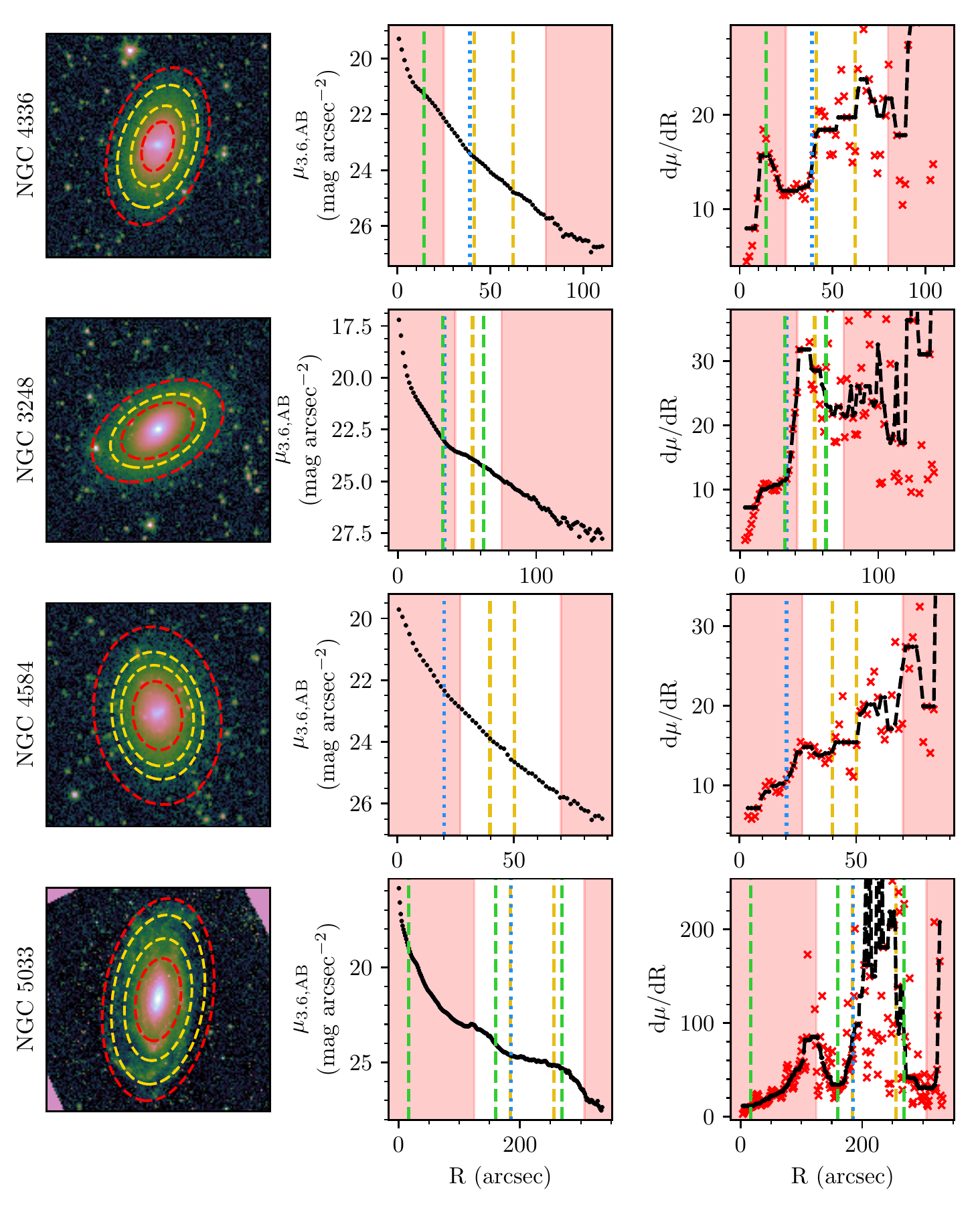}
  \caption{Demonstration of inner radius choice.  The left column
    shows 3.6$\mu$m images of the demonstration galaxies, with the
    inner and outer radius fitting boundaries shown as red
      dashed ellipses and break radii as yellow ellipses.  The middle
    column shows these galaxies' 3.6$\mu$m surface brightness
      profiles.  Red shaded regions denote the radial regions
      excluded from our break-finding procedure.  Green dashed lines
    show the radii of inner rings, lenses, etc. used in
    determining the inner radius boundaries (see text), as well
      as outer rings, lenses etc. used for aiding in break
      classification.  Yellow dashed lines mark break radii from our
      analysis, while blue dotted lines mark break radii from L16
      (break classifications are discussed in more detail below).
    Finally, the right column shows the local slope of the surface
    brightness profile vs. radius; vertical lines are the
    same as in the middle column, and the red points show the local
    slope profile before median smoothing (see text).  From the
      top row to the bottom, we demonstrate avoidance of: an inner
      ring; an inner lens; a non-exponential inner region; and an oval
      distortion.  \label{fig:inrad}}
\end{figure*}

To overcome some of the uncertainties in break classification, we have
developed a new approach to identify breaks.  First, we define inner
radius boundaries by excluding from our analysis those regions
satisfying the following criteria:

\begin{enumerate}
  
  \item The region falls inside of the bar, inner ring, inner
    pseudoring, inner lens, or inner spiral arm radius, as measured by
    \citet{herreraendoqui15}, if available (most inner spiral arm
      radii were not available, hence were estimated by eye).
    
  \item The region is not well-characterized by an exponential
    profile.
    
\end{enumerate}

These conditions avoid regions of the disk in which the shape of the
surface brightness profile is evidently dominated by bars and related
internal structures \citep[$\theta$- or $\phi$-shaped regions,
  e.g.;][]{curtis18, buta96}, as well as non-exponential structures
such as classical bulges.  Bars are known to redistribute mass within
spiral galaxies \citep[e.g.,][]{hohl71, athanassoula02, debattista06,
  binney08, minchev11, athanassoula13}, thereby altering the surface
brightness profile of the host \citep[e.g.,][]{pohlen06, erwin08,
  munozmateos13, laine14, diazgarcia16a}.  We thus choose to focus on
regions of the disk that are more likely to be influenced by external
factors.  We use the outer radius boundaries from L16, defined
  as the radius at which the surface brightness profile alters by
$\pm$0.2 magnitudes with $\pm1 \sigma_{\rm sky}$ perturbations to the
sky subtraction \citep[as][]{pohlen06}, where $\sigma_{\rm sky}$
is the standard deviation of the local background from
  \citet{salo15}.  We provide a few illustrative examples in
Fig.~\ref{fig:inrad} using four galaxies' surface brightness
and local slope profiles, plotted against radius\footnote{For
    rings, lenses, etc., and for our $b/a$ profiles, we use the
    features' or isophotes' on-sky semi-major axis lengths as their
    radius.}.

Additionally, we excluded similar regions near oval distortions
because of the way in which oval distortions act as bars
\citep[e.g.,][]{kormendy79, jogee02, kormendy04, trujillo09}.  As
  a rough preliminary selection criterion, we identified oval galaxy
candidates as those galaxies with inner isophotes showing deprojected
axial ratios of $b/a \lesssim$0.8. This is a $\sim$4--5$\sigma$
  deviation from circularity given the typical errors on the
  ellipticity \citep[e.g.,][]{munozmateos15} in the outer isophotes
  from which the axial ratio and position angle used for deprojection
  were drawn.  We then narrowed down this candidate list
through visual inspection, e.g. seeking out outer spiral arms emerging
from the ends of the oval, resulting in 23 oval disk candidates.
However, in most cases this oval classification was ambiguous --- for
example, the change in axial ratio between the inner and outer
isophotes was either very slight or very gradual --- therefore in only
5 of 23 cases did we take this oval classification into consideration
when defining our inner radius limits.  For these 5 galaxies (e.g.,
NGC 5033, shown in Fig.~\ref{fig:inrad}), we chose to avoid the oval
regions for both consistency (under the assumption they act as bars)
and for simplicity, as the surface brightness profiles of such
galaxies are extremely complex, showing upward of 5 significant
changes in slope.

\begin{figure*}
  \centering
  \includegraphics[scale=1.0]{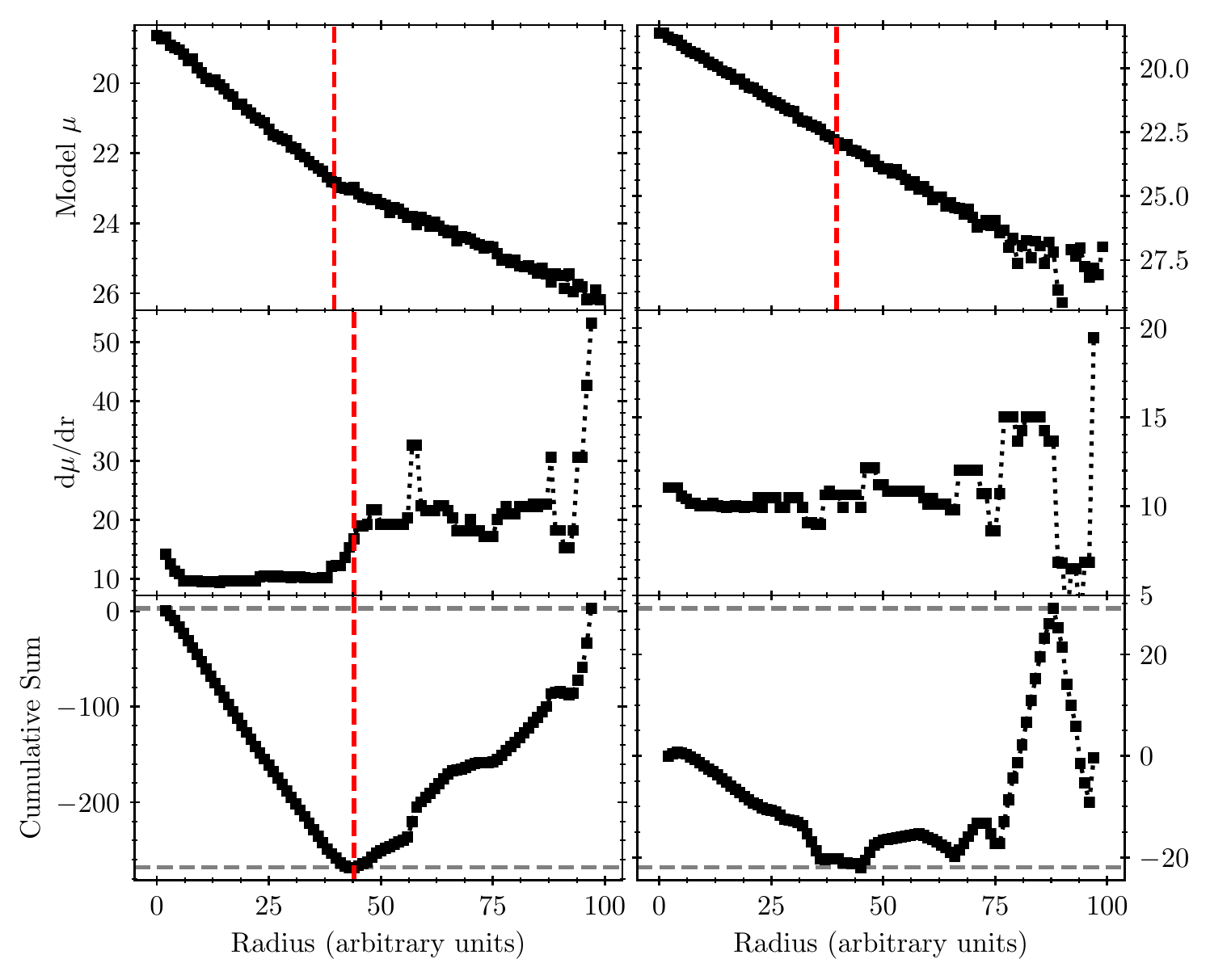}
  \caption{Demonstration of break-finding algorithm.  {\bf Top:} two
    idealized model Type~III surface brightness profiles, with Poisson
    and simulated sky noise added.  Both profiles use the same inner
    disk slope; the left profile contains a break with h$_{\rm
      in}/$h$_{\rm out} = 0.5$, while the right profile contains a break
    with h$_{\rm in}/$h$_{\rm out} = 0.9$.  Red dashed lines mark the
    break radius. {\bf Middle:} the local slope profiles of the model
    surface brightness profiles.  The red dashed line in the left
    profile marks the break radius as determined by our break-finding
    algorithm.  No significant break was found for the right profile.
    {\bf Bottom:} the cumulative sums, as defined in Equation
    \ref{eq:cusum}, with the break radius of the left profile again
    marked with a red dashed line.  Gray dashed lines mark the
      locations of the maximum and minimum values of CS (in the text,
      max(CS) and min(CS)).  Note that the minimum CS of the right
    profile occurs at the same location as the left profile; the
    break, however, was not found to be significant and is therefore
    not shown.  \label{fig:cusum}}
\end{figure*}

Once we defined the inner radius boundary, we measured the local slope
of the surface brightness profile as a function of radius, following
\citet{pohlen06}, i.e., using the four adjacent points on the
curve at each radial bin, but smoothing the resulting slope profile
with a median kernel with a width of 10\% of the full extent of the
  array rather than 5\%.  We used four adjacent points to measure
slope in order to balance the level of noise with resolution; a wider
range tends to both shift the break location and wash out subtler
breaks, for example.  Likewise, we chose this kernel width to balance
resolution with the often small angular size of the sample galaxies,
the smallest of which contain only $\sim$30 radial bins (e.g.,
ESO~402-30).  These choices influence only the subtlest breaks, and so
have minimal impact on our conclusions.

To identify break radii, we used a statistical method called
change-point analysis \citep[reproduced in Python by the authors
  following the online technical description]{taylor00}.  In brief,
this method searches for significant changes in the mean of a
time-series (or similar profile) using the cumulative sum
\citep[hereafter, CS;][]{hinkley71}.  The CS is defined for a variable
$x$ with sample size $N$ as:
\begin{equation} \label{eq:cusum}
CS_{0} = 0,\textrm{ } CS_{N}= \sum_{j=1}^{N}(x_{j} - \bar{x})
\end{equation}
where $\bar{x}$ is the mean of $x$.  Hence, CS is
actually the cumulative sum of the difference from the mean.

We show two examples of the break-finding method for idealized model
Type~III disks in Fig.~\ref{fig:cusum}, created as the sum of two
exponentials \citep[e.g.,][with $\alpha$=2]{erwin08, laine14} with
Poisson noise and artifical sky noise added to the flux at each
radius.  The left panels show a successful application of the
technique on a fairly strong disk break (with the ratio of the
  inner to outer disk scale lengths h$_{\rm in}/$h$_{\rm out} =
0.5$), while the right profile shows an example of a weak break buried
in noise, where the algorithm fails (h$_{\rm in}/$h$_{\rm out} = 0.9$,
in this case).  In the successful case in the left panels, because the
innermost slope is consistently lower than the mean slope (roughly the
average of the inner and outer slopes), the CS curve continually
decreases until the change point; beyond the change point, the slope
remains consistently higher than the mean, and the CS curve increases.
The minimum of this profile therefore marks the break radius (the red
dashed lines in Fig.~\ref{fig:cusum}).  By counterexample, the CS
curve of a Type~II break galaxy would be inverted, and the change
point would occur at the maximum.

We note that the agreement between the break location as defined by
the algorithm with the true break location depends on the smoothness
of the break.  A least-squares fit (such as that done by Laine et
al. 2014, L16) should provide more accurate break locations.
That said, given that we find as many as three breaks in many galaxies
in our re-analysis, the complexity of such profiles runs the risk of
over-fitting the curves, therefore we choose simply to identify
  the points where the slope changes rather than fit a multi-parameter
  functional form.  Because we are more concerned with the break
  classifications themselves than with the precise locations of the
  breaks, this choice does not alter any of the scientific
conclusions in our paper.

We ran this break-finding algorithm three times per galaxy.  The
first locates the global minimum of the CS profile; we then split the
profile at the identified break radius to locate additional
significant changes in slope shortward and longward of the initial
change point.  Each galaxy thus was classified with a maximum of three
disk breaks.  This choice to split the profiles three times allowed us
to identify locally important changes in slope in addition to global
changes, thereby skirting the assumption that any given galaxy be
primarily defined by just one disk break.  In a few cases (NGC~4041,
NGC5963, etc.) more than three significant disk breaks were
identifiable through further splitting of the profile; for the sake of
simplicity, however, we included only the three most significant
breaks in the final classification.

To test the significance of each break, we employed a bootstrapping
test following \citet{hinkley87, taylor00}.  For each
  galaxy, we recorded the first identified break's strength
(defined as $CS_{\rm diff}= $max$(CS) - $min$(CS)$). We then randomly
reordered the galaxy's radial slope profile, rederiving and
recording the new location of the CS minimum and the new reordered
break strength a total of $10^{5}$ times.  If the break is
significant, more often than not the break strength in the reordered
profile will fall below that of the real profile.  Therefore, we
defined as significant those breaks for which $CS_{\rm diff}$ of the
reordered CS profiles was less than $CS_{\rm diff}$ of the original
profile in $\geq$95\% of re-samplings.  We then repeated this
  process for each additional break identified in the galaxy.  We
  applied these significance tests to all breaks in all galaxies in the
  full sample.  This process reduces the number of false positives
caused by e.g. very minor or very localized changes of slope.

Next, to test the break resilience to the sky uncertainty, we perturbed
the surface brightness profiles by adding and subtracting the
1$\sigma$ flux uncertainty in each radial bin $i$, defined as
\begin{equation} \label{eq:sky}
\sigma_{i}^{2} = \sigma_{\rm sky}^{2} + \sigma_{\textrm{Poisson}, i}^{2}
\end{equation}
where $\sigma_{\rm sky}$ is the RMS in the background sky \citep[using
  the values given by][normalized by $\sqrt{N}$, with $N$ the number
  of pixels in each radial bin]{salo15} and $\sigma_{\textrm{Poisson},
  i}$ is the Poisson uncertainty in the flux in bin $i$.  We then
repeat the break-finding analysis and bootstrap resampling procedure,
rejecting all breaks if they fall below our 95\% bootstrapping
significance cutoff with either perturbation.  This minimizes the
inclusion of potentially spurious breaks resulting from improper
background subtraction.

\subsection{Break classification scheme}

Our break-finding algorithm identified many new breaks across the
galaxies in our sample compared to those previously identified by L16.
Therefore, at the risk of adding complexity to an already complex
subject, we adopt a new disk break classification scheme for our
study, building from previous schemes \citep{pohlen06, erwin08,
  laine14}.  For reference, we provide a summary of the break types in
Table \ref{tab:bclass}.  We also provide detailed justifications
  for all break classifications of all our sample galaxies, including:
  written justifications; radial profiles of surface brightness, local
  slope of surface brightness, projected and deprojected axial ratio
  and position angle, as well as $m=1$ and $m=2$ Fourier mode
  amplitude; and images, including 3.6$\mu$m, $g$-band (as well as
  unsharp-masked versions of each), deprojected 3.6$\mu$m images, and
  3.6$\mu$m images masked to outline bright structures found in the
  disk.  We used these profiles and images to aid in our break
  classifications, and make them available both through the
    Centre de Donn\'{e}es astronomiques de Strasbourg (CDS) and at the
    following URL:
    \url{https://www.oulu.fi/astronomy/S4G_TYPE3_DISC_BREAKS/breaks.html}

\begin{table*}
  \caption{Break Classification Summary}
  \label{tab:bclass}
  \centering
  \begin{tabular}{ll}
    \hline\hline
    Type 0: & Not exponential \\
    Type I: & No significant disk break \\
    Type IId: & Down-bending break at spirals, lenses, rings, etc. \\
    Type IIa: & Down-bending break at asymmetric feature (e.g., tidal stream) \\
    Type IIIa: & Up-bending break at asymmetric feature \\
    Type IIId: & Up-bending break marking symmetric outer disklike features, e.g., spiral arms or HII regions \\
    Type IIIs: & Up-bending break marking outer spheroidal component \\
\hline
  \end{tabular}
  \tablefoot{The nomenclature Type~I, Type~IIId, and Type~IIIs
      are borrowed from previous studies, and are defined similarly
      \citep{freeman70, pohlen06, erwin08}.  Type~IIa breaks are akin
      to Type~II-AB from \citet{pohlen06, erwin08}.  Type~IId breaks
      include all Type~II break subtypes from \citet{pohlen06,
        erwin08, laine14}, L16 except those inside the bar radius.}
\end{table*}

\begin{figure*}
  \centering
  \includegraphics[scale=01.0]{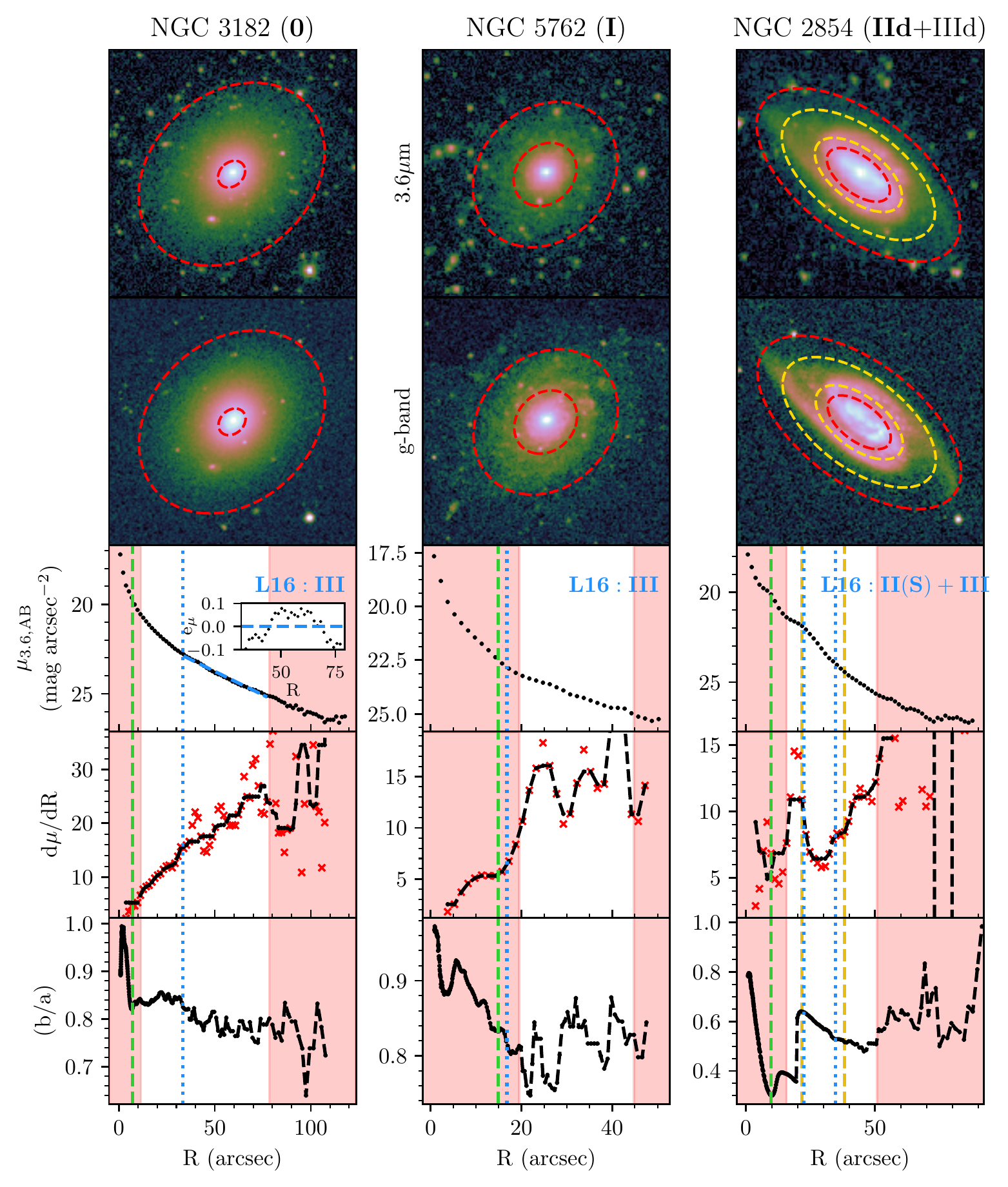}
  \caption{Examples of our break classification scheme.  We show
    here galaxies of Type~0 (left), Type~I (middle), and a
      Type~IId break (right, innermost break).  These are
    reclassifications of galaxies with Type~III breaks identified by
    L16. Bold text indicates the break being demonstrated, and
    the original break classifications are provided in the third
      row panels.  Ellipses are as in Fig.~\ref{fig:inrad}.  The
    third row shows the 3.6$\mu$m surface brightness profile (AB
    magnitudes) vs. radius.  The inset in the left panel
      shows the residuals to a linear fit (shown as the dashed blue
      line) to the surface brightness profile beyond the break radius,
      demonstrating its non-exponential nature.  The fourth row shows
    the local slope of the surface brightness profile vs. radius.  The
    fifth row shows the galaxy's on sky axial ratio vs.
      radius.  Vertical lines and shaded regions are as in
      Fig.~\ref{fig:inrad}.  See the text for specific details on
    each galaxy's break classification. \label{fig:btypes1}}
\end{figure*}

\begin{figure*}
  \centering
  \includegraphics[scale=1.0]{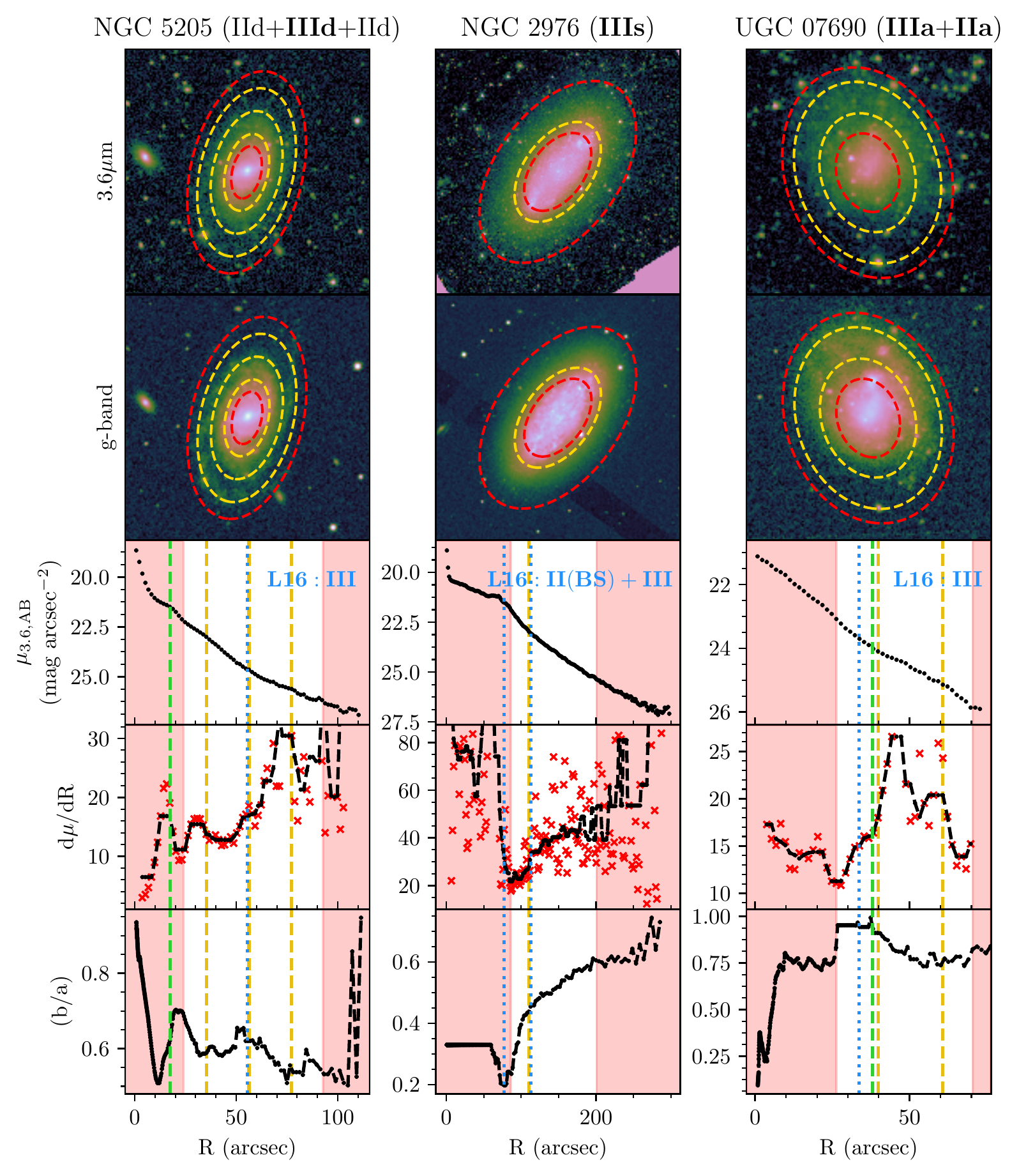}
  \caption{As Fig.~\ref{fig:btypes1}, now demonstrating a Type~IIId
    break (left column), a Type~IIIs break (middle column), and Type
    IIIa and IIa breaks (right column), each again reclassifications
    of Type~III break hosts as identified by L16.  Ellipses,
      vertical lines, and shaded regions are as in
      Figs. \ref{fig:inrad} and
      \ref{fig:btypes1}.  \label{fig:btypes2}}
\end{figure*}

\subsubsection{Type~0 galaxies}
In several cases, we found that the galaxies in our sample,
  though morphologically late-type, are better fit by a S\'{e}rsic
  profile with a S\'{e}rsic index $n > 1$.  We therefore define these
  galaxies as Type~0, to denote that they are not exponential disks.
  Because they are not exponential, the automated $\chi^{2}$
minimization procedure used by L16 --- which assumes the profile is
exponential --- always finds a Type~III profile as the best fit,
albeit with a more or less arbitrary break radius given that the
exponential slope of this profile continually rises.  We include
  these galaxies only because they were classified as Type~III disks
  by L16; clearly, future disk break studies should only be concerned
  with exponential disks.  This is a rare problem: we classify only 6
galaxies as Type~0.

We identified such cases first through visual inspection
of the local slope profiles, then by fitting the surface
  brightness profiles as exponentials and inspecting the residuals to
  the fits, which in non-exponential cases are highly non-normal.  We
show an example of this in the left column of
Fig.~\ref{fig:btypes1}.  In some cases, while the bulk of the profile is
perhaps best parametrized with a S\'{e}rsic index $n > 1$, plateaus in
slope can be found in certain radial bins.  If any plateaus in slope
could be identified we chose not to classify the galaxy as Type~0, as
these exponential regions within the galaxy potentially trace light
from a more disklike component.

\subsubsection{Type~I disks}
We define Type~I disks following \citet{freeman70}, as disks with
surface brightness profiles described well by a single exponential
slope.  An example of a Type~I profile is shown in the middle column of
Fig.~\ref{fig:btypes1}.  In this example, the Type~I designation is
appropriate only through avoidance of the inner lens structure
\citep[with a radius marked by the green dashed line;][]{buta15,
  herreraendoqui15}, which is evidently the cause of the Type~III
break identified by L16 (marked by the blue dotted line).
In our reclassifications, we identify only 10 galaxies as Type~I,
always through avoidance of such inner structures.

\subsubsection{Type~II breaks}

Because we avoid all regions of the disk near the bar, and because our
primary interest is in Type~III disk breaks, we opt for more general
Type~II subclassifications than those employed by \citet{pohlen06}.
First, we classify as Type~IId any Type~II breaks that appear
associated with spiral structure, outer lenses, and outer rings
(structures associated with the disk).  Second, we classify as Type
IIa any Type~II breaks that appear associated with tidal streams or
other asymmetric features, such as single spiral arm modes 
  \citep[akin to the Type~II-AB classification from][]{pohlen06}.
Such features are often visually obvious, however in ambiguous cases
we examined the galaxies' $m=1$ Fourier mode amplitude profiles,
seeking out $m=1$ amplitudes in excess of $\sim$0.2 at or just beyond
the break radius.  In only one case could we not identify any clear
feature associated with the Type~II break, hence we classify this
break simply as Type~II, with no suffix.  This classification scheme
is a simplification of the scheme adopted by L16, in which we ignore
the differences between breaks associated with outer lenses, rings,
pseudorings, and spiral structure.

We show an example Type~IId break in the right column of
Fig.~\ref{fig:btypes1}; the white ellipse shows the IId break radius
falling almost perfectly along the ridge of the galaxy's double spiral
arms.  In this case, however, the slope flattens into a Type~III break
afterward when the isophotes encounter these arms a second time due to
their loose winding.  We show an example Type~IIa break in the
rightmost column of Fig.~\ref{fig:btypes2}, in which the break is
associated with a single spiral arm in the galaxy's north (the onset
of which is marked, again, by a Type~III break).

\subsubsection{Type~III breaks}

We first adopt the two Type~III disk break subclassifications from
\citet{pohlen06}: Type~IIId and IIIs.  The former refers to upbending
disk breaks that appear associated with enhanced star formation or
spiral structure in the outer disk, the latter to a spheroidal
component with rounder isophotes surrounding the host \citep{erwin08}.

The right column of Fig.~\ref{fig:btypes1} and the left column of
Fig.~\ref{fig:btypes2} demonstrate Type~IIId breaks: in both galaxies,
a Type~IIId break occurs when the photometry aperture first encounters
a set of double spiral arms in the disk outskirts.  In invoking this
classification, we assume such features are not tidal in nature,
though this may be inaccurate in some cases.

Type~IIIs breaks are identifiable through symmetric outer isophotes
that gradually increase in axial ratio $b/a$ with increasing radius.
For low-inclination galaxies, it is unclear if this enhanced roundness
is truly related to a vertical thickening; however, in the
  absence of information regarding the vertical thickness we simply
apply the Type~IIIs label to all galaxies demonstrating this behavior,
regardless of inclination.  A clear example of a Type~IIIs break is
shown in the middle column of Fig.~\ref{fig:btypes2}; note the
continual increase in projected axial ratio beginning at roughly
  R$\sim$70\arcsec \ despite the relatively flat local slope profile.

Additionally, we introduce a new subtype: Type~IIIa.  Akin to the Type
IIa breaks, this refers to Type~III breaks apparently associated with
asymmetric features and high $m=1$ Fourier amplitudes (again, in
excess of $\sim$0.2).  This includes galaxies with offset bars and/or
single spiral arm modes, clearly tidally disturbed galaxies, and
galaxies surrounded by disorganized tidal debris, which previous
studies have shown can level out the surface brightness profiles where
they occur \citep[e.g., NGC~4319, M87, and NGC~474
  from][respectively, although in M87 this may also be due to
  contamination from Galactic cirrus]{erwin05, janowiecki10, laine14}.
The reason for this resides in the azimuthal averaging process: when
the photometry aperture encounters a tidal feature that is elevated in
surface brightness above the disk at the same radius, if the aperture
traces the feature over an extended azimuth, the mean surface
brightness in that aperture will reflect that of the tidal feature and
not the underlying disk.  We show an example of this break type in the
right column of Fig.~\ref{fig:btypes2}, where the first break radius
in this galaxy clearly occurs at the onset of the galaxy's offset
outer pseudoring, while the southern half of the photometry aperture
tracing the inner disk more or less reaches the background limit at
this same radius.  In this case, we also identified a Type~II break at
the ridge of the pseudoring.

In a handful of cases, the origin of the Type~III break was unclear.
Therefore, as with Type~II breaks, we classified these galaxies as Type
III, without a suffix.

\subsection{Comparison with previous studies}

We briefly compare here our break classifications with those of the
following previous studies: \citet{erwin05, pohlen06, erwin08,
  gutierrez11, erwin12}.  Our sample overlaps with these previous
studies by only 23 galaxies.  Of these 23, we find rough matches
between breaks we have identified and those identified in these
previous studies in only 10 galaxies.  In all cases in which we find
matching breaks, the directions of the breaks (Type~II or Type~III)
agree, though in only 2 cases (NGC~2967 and NGC~3982) do we give the
same subclassification to the break in question (both of Type~IIId).
Most often, we classify as Type~IIIa what had previously been
classified as Type~IIId (5 of 10 cases).  One notable overlap between
our sample and these previous studies is NGC~7743 (discussed in detail
below), which \citet{erwin08} classified as Type I despite its rather
complex profile.  In general, then, we find poor agreement with
previous studies.

L16 found similarly poor agreement, identifying matching breaks in
fewer than $2/3$ of the galaxies they compared across studies.  They
concluded that the primary source of confusion was likely the lack of
definite criteria for defining disk breaks (leading to subjective
judgment), and differences among studies' surface brightness limits.
While we have no control over survey surface brightness limits, our
algorithm does avoid the subjectivity inherent in break detection, and
therefore should be significantly more reproducible.  Regardless, our
comparison here and that of L16 illustrates the need for greater
consistency across disk break studies.

\section{Results}

\subsection{Statistics by Break type}

\begin{table}
  \caption{Frequency of Break types}
  \label{tab:bfracs}
  \centering
  \begin{tabular}{lcc}
    \hline\hline
    & Total $\#$ & $\#$ as Outermost \\
    \hline
    Type II & 1 ($<$0.01) & 1 ($<$0.01) \\
    Type IId & 74 (0.24) & 25 (0.14) \\
    Type IIa & 27 (0.09) & 17 (0.10) \\
    Type III & 3 (0.01)  & 3 (0.02) \\
    Type IIId & 96 (0.31) & 31 (0.18) \\
    Type IIIs & 24 (0.08) & 21 (0.12) \\
    Type IIIa & 89 (0.28) & 61 (0.35) \\
    All & 314 & 175 \\
    \hline
  \end{tabular}
\tablefoot{Reclassifications of breaks in the Type~III break hosts
  identified by L16.  We provide both the total number of breaks by
  break type we identified at all radii, and the total number of
  breaks by break type that constitute the outermost break of their
  host galaxy.  The final row gives the total of each column.
 Numbers in parentheses give the fraction of the total.  This excludes
galaxies with no breaks (9\% of all galaxies), hence the total
fraction in the second column is only 91\%.}
\end{table}

\begin{table}
  \caption{General Break Statistics}
  \label{tab:bstats}
  \centering
  \begin{tabular}{lc}
    \hline\hline
    Num. galaxies w/1 break & 40 (0.23) \\
    Num. galaxies w/2 breaks & 82 (0.47) \\
    Num. galaxies w/3 breaks & 37 (0.21) \\
    Num. galaxies of Type I & 10 (0.06) \\
    Num. galaxies of Type 0 & 6 (0.03) \\
    \hline
  \end{tabular}
  \tablefoot{Total number of galaxies with 1, 2, 3, or no breaks, in
    our reclassifications of Type~III break hosts identified by L16.
    Numbers in parentheses give the fraction of the total.}
\end{table}

\begin{table*}
    \caption{Break Radii and Classifications}
    \label{tab:breaks}
    \centering
    \begin{tabular}{cccccccccc}
      \hline\hline
      (1) & (2) & (3) & (4) & (5) & (6) & (7) & (8) & ... & (18) \\
Galaxy & B$_{\rm type}$ & R$_{\rm in}$ & R$_{\rm out}$ & R$_{\rm br,1}$ & R$_{\rm br,2}$ & R$_{\rm br,3}$ & $\mu_{\rm br,1}$ &  ... & $\mu_{0,4}$ \\
  & & [\arcsec] & [\arcsec] & [\arcsec] & [\arcsec] & [\arcsec] & [AB, 3.6$\mu$m] &  & [AB, 3.6$\mu$m] \\
\hline
NGC3155 & IIs+IIId+IIId & 8.28 & 46.10 & 17.25 & 27.75 & 38.25 & 22.17 & ... & 22.74 \\
NGC3177 & IIId+IIIa+IIs & 17.32 & 51.50 & 23.25 & 32.25 & 39.75 & 22.10 & ... & 19.65 \\
NGC3182 & 0 & 11.43 & 78.40 & -- & -- & -- & -- & ... & -- \\
NGC3248 & IIs & 41.41 & 75.00 & 54.00 & -- & -- & 23.89 & ... & -- \\
NGC3259 & IIId+IIIa & 11.31 & 60.70 & 24.75 & 42.75 & -- & 22.44 & ... & -- \\
NGC3274 & IIs+IIIa & 11.29 & 89.60 & 41.25 & 53.25 & -- & 24.13 & ... & -- \\
\hline
    \end{tabular}
    \tablefoot{Sample of the table which will be available at the CDS.
      Data that will be available is as follows: (1) galaxy name as
      listed in S$^{4}$G catalog, (2) break classification, (3) inner
      radius boundary, (4) outer radius boundary, (5)--(7) first
      through third break radii,
      (8)--(10) 3.6 $\mu$m surface brightness at first through third
      break radii, (11)--(14) scale lengths of linear fits between
      break radii and between break radii and radial boundaries,
      (15)--(18) central surface brightness of linear fits between
      break radii or between break radii and radial boundaries.}
\end{table*}

Tables \ref{tab:bfracs} and \ref{tab:bstats} provide general
  statistics of the breaks across our sample.  Table \ref{tab:breaks}
  provides the details of the fitting parameters; the full version
will be available through CDS.

In total, among 175 galaxies, we identified 314 significant disk
breaks.  Of these, 100 were not identified in the previous analysis by
L16, suggesting that were we to run this analysis on the remaining L16
sample (Type~I and II disks), the break statistics of that sample
would change significantly as well (though this is beyond the scope of
the current paper).  Our classifications broadly agreed with
  those of L16 in $\sim$27\% of galaxies (ignoring
  subclassifications).  We found a more complex profile than that of
  L16 in $\sim$56\% of cases.  In the remainder of cases, we found
  fewer breaks than L16, either through rejection of their innermost
  break using our revised inner radius threshold criteria, or through
  rejection of their outermost break through our improved masking
  (though the latter occurred in only two cases).  Of the current
sample, most galaxies contained two breaks ($\sim$47\%), followed by
those with only one break ($\sim$23\%) and those with three breaks
($\sim$21\%).  In 10 of the galaxies we found no break ($\sim$6\%),
always as a result of our choice to exclude the disk regions inside of
inner rings, lenses, bars, ovals, and other such structures based on
the classifications by \citet{buta15}.  An additional 6 galaxies
($\sim$3\%) lacked exponential profiles and so were classified Type~0.
The most common break types in our sample are IIId (31\% of all
identified breaks), IIIa (28\%), and IId (24\%).  Aside from Types II
and III (those breaks with no obvious cause), the least common break
type in our sample is IIIs (8\% of all breaks).

Evidently, in those galaxies in which we do identify breaks, our
break-finding algorithm is far more liberal than those of previous
studies, which most often assign galaxies only one break
  \citep[e.g.,][who give hybrid classifications to only 10\% of their
    sample galaxies]{pohlen06}.  However, we argue that our
break-finding algorithm relies on fewer assumptions than those of
previous studies by minimizing the need for human judgment: as long as
a change of slope is significant and lasting, relative to its
surroundings, it is counted, regardless of its origin or strength.
One consequence of this was the identification of 26 outer Type~II
breaks that were missed by the previous analysis of this data (L16).
Many of these were subtle in 3.6$\mu$m, therefore whenever possible we
confirmed these Type~II classifications through examination of optical
images using either SDSS, LT imaging from \citet{knapen14}, or from
searches for deep amateur or professional photography published online
\citep[for example, the Carnegie-Irvine Galaxy Survey;][]{ho11}.  In
all cases with such imaging available we found that the breaks
identified by our algorithm aligned with spiral structure more easily
visible in optical images (an example of this can be seen in the left
column of Fig.~\ref{fig:btypes2}).

Many of the breaks we identified, of course, are followed by
additional breaks and so should not be used to characterize the
outermost disk, which is the focus of this study.  We therefore focus
only on the outermost break in each galaxy, that which is most
susceptible to environmental influences.  In Section 5.1, we
  discuss the influence of the S$^{4}$G's surface brightness limit on
  this choice, as well as the influence of our limit of three breaks
  per galaxy.

Of all the sample galaxies with disk breaks, $\sim$45\% of the
outermost breaks seem related to tidal debris or other asymmetries
(IIIa and IIa), and $\sim$15\% are down-bending and related to spiral
arms or similar features (IId).  About 9\% of the galaxies show no
breaks outside of those caused by internal structures, e.g.,
  inner rings and lenses.  Only $\sim$30\% of our sample galaxies are
therefore best characterized as more or less symmetric disks with a
persistent excess of light in their outskirts (IIId and IIIs).  Given
that our sample contains only galaxies previously classified as having
Type~III breaks, and given that Type~III breaks characterize
$\sim$36\% of all disk galaxies in the Local Universe
\citep[e.g.,][with inter-study spread of $\sim \pm$10\%]{pohlen06,
  erwin08, gutierrez11, maltby12, laine14, pranger17, wang18}, this
suggests that such galaxies are quite rare in the general population
($\sim$30\% of $\sim$36\%, or $\sim$11\% of all Local Universe
galaxies).

\subsection{Color Profiles}

\begin{figure*}
  \centering
  \includegraphics[scale=1.0]{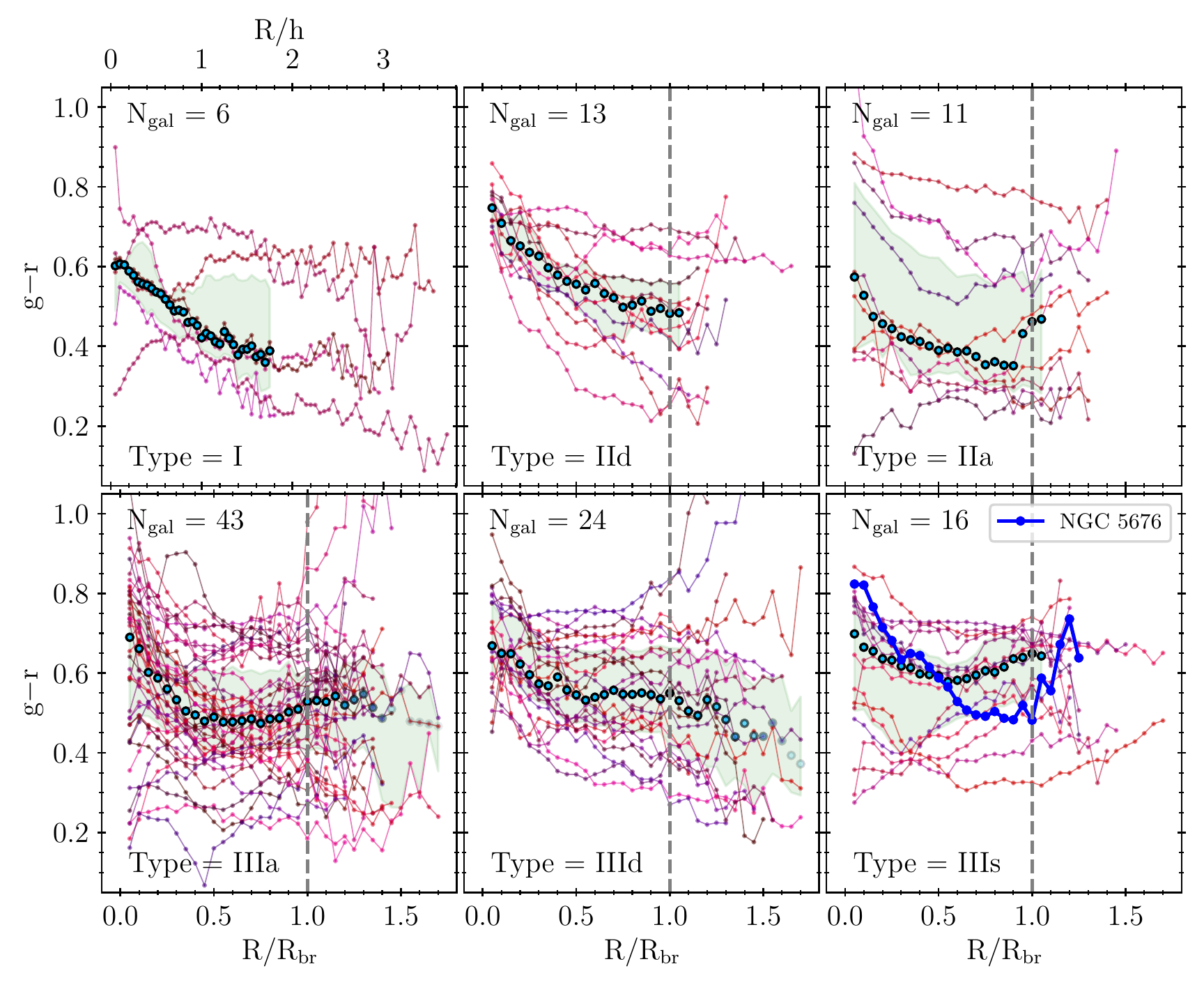}
  \caption{$g-r$ color profiles of sample galaxies, with radii
    normalized by the radius of the outermost break.  In the case of
    Type~I galaxies, the profiles are normalized instead by the disk
    scale length.  Curves and points in shades of red through
      purple denote individual galaxy profiles, with the total number
      of profiles in each panel given as N$_{\rm gal}$.  Light blue
      points denote the median of all profiles at each radial bin,
      with the interquartile range (IQR) shown as the light green shaded
      region.  For panels with N$_{\rm gal} < 20$, we show the median
      values and IQRs only where all galaxies contribute; for those
      with N$_{\rm gal} > 20$, marker transparency is calculated as
      the number of contributing galaxies at that radial bin divided
      by N$_{\rm gal}$.  The blue curve in the lower right panel shows the
      color profile of NGC~5676, the only Type~IIIs galaxy in our
      sample with an extended blueward color gradient (see
      text).  \label{fig:cprofs}}
\end{figure*}

Following \citet{bakos08}, we show in Fig.~\ref{fig:cprofs} median
$g-r$ color profiles of the galaxies in our sample, separated by
outermost break type.  Small multi-colored points in each panel
show individual galaxy profiles, normalized by the outermost break
radius R$_{\rm br}$ (and the disk scale length, in the case of Type~I
galaxies), while large light blue points show the running
median, with the interquartile range (IQR) for all galaxies shown
  as the green shaded region.  We truncate these median profiles for
  panels with fewer than 20 galaxies to the maximum radius of the
  shortest profile, as beyond this point the median values at each
  radial bin are misleading given the very small number of profiles
  contributing.  For panels with more than 20 galaxies, the fewer the
  number of contributing galaxies at each radial bin, the more
  transparent the point.  We note that for all break types, $g-i$
color profiles show the same behavior as these $g-r$ profiles,
despite significantly higher background noise in the $i$-band images.

Given the lack of galaxies contributing to the median profiles
  for most break subclassifications, it is difficult to assess common
  trends.  Type~I profiles, for example, show a wide variety of
  behaviors: two galaxies trend continually blue, another trends blue
  but flattens beyond $\sim$2$h$, one trends blue only beyond
  $\sim$2$h$, and one more is roughly flat throughout.  The median
  profile is therefore not reflective of the individual galaxy
  behaviors.  However, Type~IId galaxies show more consistent
  behavior, with the majority trending blue up to the break radius (as
  denoted by the median profile).  Only one Type~IId galaxy (NGC~3177)
  shows a distinct U-shaped profile akin to the average profiles of
  Type~II disks found by \citet{bakos08, zheng15}.  In our galaxies,
  most of our outermost Type~IId breaks occur at such low surface
  brightness that color information is not available much beyond
  R$_{\rm br}$.  Similarly, while the median profile for Type~IIa
  galaxies shows a rough U-shape, the individual galaxy profiles show
  primarily either steadily blueward gradients, flat blue profiles, or
  blue inner disks with redward gradients in the outer disk.  In the
  latter cases, the outermost redward gradients begin at roughly the
  break radius.  Again, in only one case, NGC~5774, does the profile
  show a clear U-shape with a turnaround at the break radius.  While
  it is encouraging that in several cases these downbending breaks,
  despite occurring at large radii and low surface brightness, appear
  to mark a star formation threshold like their higher surface
  brightness counterparts in other studies \citep{bakos08, zheng15},
  deeper multiband imaging is clearly required to determine whether or
  not this is true for most of the galaxies in our sample.

Trends are more robust for Type~IIIa and IIId galaxies given the
  larger sample sizes.  Type~IIIa disks show on average initial
  blueward gradients followed by fairly flat profiles averaging $g-r
  \sim$0.5.  The radial extent of this bluer outer disk relative to
  the disk break in Type~IIIa galaxies may be greater than that of
  Type~II galaxies simply by definition: for example, the
    breaks in Type~IIa disks mark the peak surface brightness of
    asymmetric features, while the breaks in IIIa disks mark the
    onset.  If such galaxies are tidally interacting, this may also
  in many cases have resulted in elevated outer disk star formation
  rates, leading to sustained blue colors on average in the outskirts
  compared to typical Type~IId galaxies (including possibly the
    Type~IId galaxies in this study, if their steady blueward
    gradients also turn around at the break radius), modulo any
  extinction from dust.

Type~IIId galaxies show decreasing color profiles on average, save a
plateau between R$/$R$_{\rm br} = 0.5$--$1.0$, with $g-r \sim 0.7$ in
their cores and $g-r \sim$0.4 in their outskirts.  This color range is
similar to the profiles of the Type~IId galaxies inside of the break
radius.  If indeed the Type~III breaks in these galaxies arise due to
extended star formation or spiral arms, one might therefore expect
many of these galaxies to eventually show a Type~IId break (and red
color upturn, potentially) in their outskirts as well, which the
S$^{4}$G and SDSS imaging is simply too shallow to locate.

Finally, the average color profile of Type~IIIs galaxies appears much
flatter than those of other types, with on average redder outer disks
($g-r\sim$ 0.6).  While the disk colors of the individual IIIs
galaxies display significant heterogeneity beyond $\sim 0.5$R$_{\rm
    br}$, inside of this radius the profiles are split into two
groups: those with $g-r \sim 0.4 \pm0.1$ (6 galaxies), and those with
$g-r \sim 0.7 \pm0.1$ (10 galaxies).  All of the galaxies with blue
inner regions eventually show redward color gradients, albeit with
different slopes and starting radii.  Only one Type~IIIs galaxy shows
an extended blueward color gradient (NGC~5676, shown as a blue
  curve in Fig.~\ref{fig:cprofs}), though even this
galaxy shows a red upturn beyond R$/$R$_{\rm br} \sim 1$.

\subsection{Environmental correlation}

Using our new break classification system, we test for correlations
between break type and the local density of galaxies.  Again, we test
this using only the outermost breaks in each galaxy, which should be the
most sensitive to the local environment.

\subsubsection{Environmental Parameters}

We compared breaks types using three measures of the local density.
First, we use \citep[as measured by][]{laine14} the Dahari parameter
$Q$ \citep{dahari84}, a measure of tidal strength, and
$\Sigma_{N}^{A}$, the projected surface number density within the
$N^{\rm th}$ nearest-neighbor distance \citep{dressler80,
  cappellari11}.  Briefly, $Q$ is defined as:
\begin{equation}\label{eq:dahari2}
  Q_{i} = \frac{(D_{\rm p}D_{\rm c})^{\gamma}}{S^{3}}
\end{equation}
where $D_{\rm p}$ and $D_{\rm c}$ are the diameters of the primary and
companion galaxies, respectively, $S$ is their projected separation,
and $\gamma = 1.5$ \citep{rubin82, dahari84, verley07, laine14}, and
$\Sigma_{N}^{A}$ is defined as:
\begin{equation}\label{eq:sigma}
  \Sigma_{N}^{A} = \log_{10}\left(\frac{N_{\rm gal}}{\pi
    R_{N}^{2}}\right)
\end{equation}
where $N_{gal}=N=3$ and $R_{N} = R_{3}$ is the projected distance to
the third nearest neighbor.  For details, we refer the reader to
  \citet{laine14}.

Finally, we also measure the surface number density within the galaxy
groups defined by \citet{kourkchi17} (hereafter, KT17), using the
  group's second turnaround radius (Equation 6 of KT17, using the $K_{s}$
  magnitude to estimate mass) to define the group surface
area.  We thus define this surface number density as:
\begin{equation}
  \Sigma_{KT17}^{A} = \log_{10}\left(\frac{N_{\rm group}}{\pi
    R_{2t}^{2}} \right)
\end{equation}
where $N_{\rm group}$ is the total number of members of the group to
which the break-hosting galaxy belongs, and $R_{2t}$ is the
group's second turnaround radius.  We use this surface
density as an independent check against $\Sigma_{3}^{A}$, which
measures density on quite local scales \citep[e.g.,][]{cappellari11}.
For details on how these groups were defined, we refer the reader
  to KT17.

\subsubsection{Statistical Tests}

Given that we are performing comparisons of the means of environmental
parameters between small sample populations (175 total galaxies split
unevenly into 7 separate categories), we require a statistical test
with more interpretive power, such that it minimizes the
false-negative (type II error) rate.  We therefore choose to perform a
Bayesian comparison of means, using the method described by
\citet{kruschke13}, dubbed BEST (for Bayesian Estimation Supersedes
the \emph{T} test).  This method is a Bayesian corollary to Student's
\emph{t} test \citep{student08}, where the means and standard
deviations are compared by their posterior distributions via Bayes'
formula:
\begin{equation}
p(\mu_{1}, \sigma_{1}, \mu_{2}, \sigma_{2}, \nu \vert D) = p(D \vert
\mu_{1}, \sigma_{1}, \mu_{2}, \sigma_{2}, \nu) \cdot p(\mu_{1},
\sigma_{1}, \mu_{2}, \sigma_{2}, \nu)
\end{equation}
for observations $D$, with $\mu_{1}, \sigma_{1}$ the mean and standard
deviation of population 1, $\mu_{2}, \sigma_{2}$ the mean and standard
deviation of population 2, and $\nu$ the number of degrees of freedom
(as well as a normalization term, not shown).  In the right-hand
  side of Equation 5, the first term is the likelihood and the second
  term is the prior (described in detail below).

\begin{figure}
  \centering
  \includegraphics[scale=1.0]{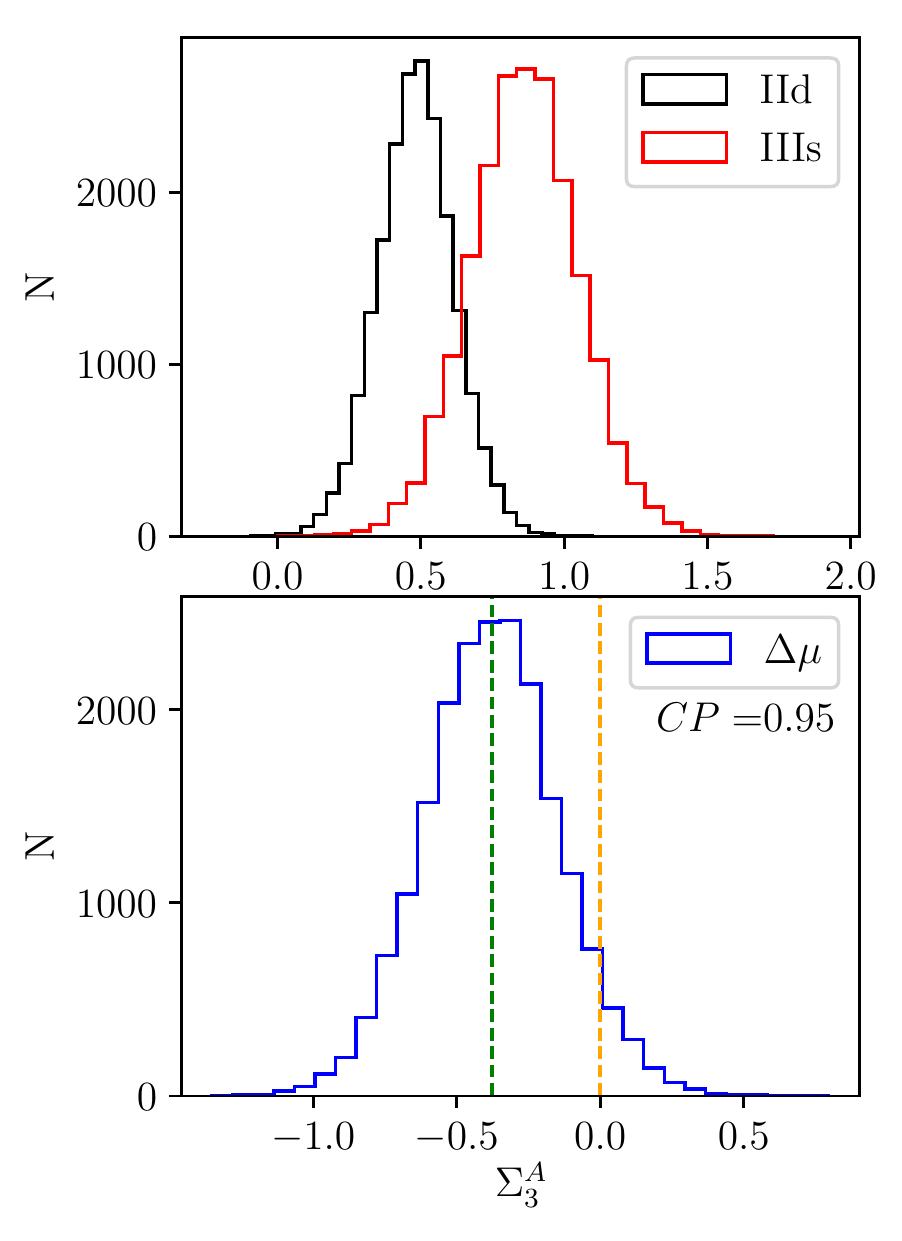}
  \caption{Demonstrating the posterior distributions of the mean
    vs. that of the difference in means, using the BEST method (see
    text).  {\bf Top:} the posterior distributions of the mean third
    nearest neighbor surface density $\Sigma_{3}^{A}$ for galaxies
    with outermost breaks of Type~IId and IIIs.  {\bf Bottom:} the
    posterior distribution of the difference in the mean
    $\Sigma_{3}^{A}$ between Type~IId and IIIs galaxies.  The green
    dashed line shows the median of this distribution, while the orange
    dashed line shows $\Delta \mu = 0.0$.  The means differ by
    $\sim$1.7$\sigma$, and likewise the posterior distribution of
    $\Delta \mu$ peaks below 0.0 with $\sim$95\%
    confidence. \label{fig:demo}}
\end{figure}

To test for differences between galaxy populations, we use this method
to compare the posterior distributions of the difference of means
$\Delta \mu \equiv \mu_{1} - \mu_{2}$ of each environmental parameter,
evaluated between all possible pairs of samples (21 comparisons for
each environmental parameter).  In each comparison, we adopt Gaussian
priors on the population means (with prior means equal to the measured
sample means), a wide uniform prior $U(0, 5\sigma_{\rm tot})$ on the
population standard deviations (with $\sigma_{\rm tot}$ the pooled standard
deviation of the environmental parameter for the full 175 galaxy
sample), and an exponential prior on $\nu$ of $\exp[-(\nu -
  1)/\lambda]$ with $\lambda = 29$, as recommended
  by \citet{kruschke13} as a balance between nearly normal distributions
($\lambda > 30$) and distributions with heavy tails ($\lambda < 30$).
We find through experimentation with model Gaussian distributions that
the results of the test are robust to the choice of priors, save that
the use of too narrow priors on the standard deviations can result in
posterior distributions of the means that are not reflective of
the input simulated data.  We assess the confidence level of each
comparison using the cumulative probability of the posterior
(hereafter, CP) up to a value of $\Delta \mu = 0.0$ (a value that
implies no difference in means).  For example, if 90\% of the
posterior distribution on $\Delta \mu$ lies below 0.0 and only 10\%
above it, we may say with $\sim$90\% confidence that $\Delta \mu <
0.0$.  A value of CP$=50$\%, by contrast, implies no difference in
populations.  We show an example in Fig.~\ref{fig:demo}, which
  implies that, with 95\% confidence, Type~IId galaxies are found in
  lower density environments than Type~IIIs galaxies, if one measures
  density using $\Sigma_{3}^{A}$.

\subsubsection{Results of Statistical Tests}

\begin{figure}
  \centering
  \includegraphics[scale=1.0]{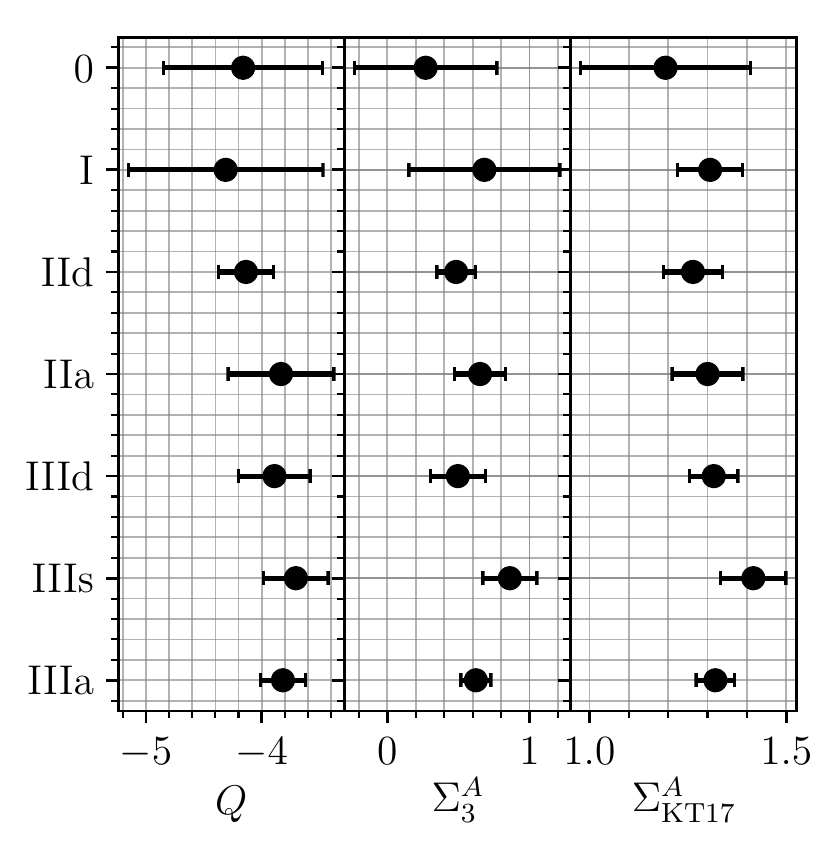}
  \caption{The means by outermost break type of three different
    environmental parameters.  Error bars are the standard deviations
    of the posterior distributions of the means, following the BEST
    method (see text).  The environmental parameters are, from left to
    right: Dahari $Q$, third nearest neighbor surface number density,
    and surface number density within the galaxy groups defined by
    KT17 (see text for full explanations).  \label{fig:scatter}}
\end{figure}

In Fig.~\ref{fig:scatter}, we show the means of all three
environmental parameters for galaxy populations separated by their
outermost break type.  The errorbars show the standard deviations of
the posterior distributions of these means, following the BEST
analysis (not the sample standard deviations).  Though intragroup
differences among these environmental parameters are small, we do find
significant trends.  Specifically, using all three environmental
parameters, we find consistently that Type~IIIs galaxies occupy the
highest density environments, and either Type~IId or 0 occupy the
lowest density environments (although the uncertainty for Type~0
galaxies is very high given the small sample size, $N=6$).
Additionally, Types IIa and IIIa typically show quite similar means,
although the standard deviations are large for Type~IIa.

In the interest of brevity, we showcase here only the strongest
trends.  Consistently, we find that Type~IId galaxies occupy lower
density environments than Type~IIIs galaxies (CPs of 87\%, 95\%, and
92\% for $Q$, $\Sigma_{3}^{A}$, and $\Sigma_{\rm KT17}^{A}$,
respectively).  This trend occurs even when using velocity limits on
$Q$ and $\Sigma_{3}^{A}$ of 500 km s$^{-1}$, albeit at lower
significance (CP $=$81\% and 78\%, respectively), suggesting it is
robust.  Indeed, Type~IIIs galaxies may occupy higher density
environments than all other break types, though at the lowest end
(IIIs compared with IIa using $Q$) the CP is only 60\%.
Comparing a combined sample of Types I, IId, IIa, IIIa, and IIId
against the IIIs sample gives CP$\sim$90\% that IIIs galaxies occupy
the densest environments.

A similarly robust trend is found regarding the Types IIa and IIIa, both of
which appear to occupy similarly dense local environments (CPs of
52\%, 56\%, and 58\% for $Q$, $\Sigma_{3}^{A}$, and $\Sigma_{\rm
  KT17}^{A}$, respectively).  With lower confidence, we also find that
Type~IId galaxies occupy lower density environments than Types IIIa
and IIa (CP $\sim$70--80\%, or $\sim$85\% using IIIa and IIa
combined), suggesting that Type~IId galaxies occupy some of the lowest
density environments in our sample, including, potentially, with
respect to IIId galaxies (CP $\sim$70\%, though the comparison with
$\Sigma_{3}^{A}$ suggests both populations occupy similar density
environments).  Finally, we find no significant trends regarding the
Type~I and 0 galaxies, though this may simply be due to their very
low sample sizes (10 and 6, respectively).

As an alternative, we also performed pairwise comparisons using the
Mann-Whitney U test \citep{mann47}.  The trends regarding Types IId
and IIIs remained insofar as tests comparing these break types
consistently yielded the lowest p-values (with, again, Type~IId
galaxies seemingly occupying the lower density environments).
However, these tests typically resulted in p-values $< 0.2$ or $<
0.1$, hence would not be considered significant under standard
significance testing criteria (especially so if one chooses to correct
for multiple comparisons).  Therefore, while these comparisons are
somewhat encouraging, these tests should be repeated using a larger
sample size in order to verify these trends.

For now, a tentative summary follows: Type~IId galaxies may occupy the
lowest density environments, followed by Types IIIa and IIa, followed
by Type~IIIs, which occupy the highest density environments.  Trends
regarding the remaining break subtypes are ambiguous.

\subsubsection{Break strengths}

\begin{figure}
  \centering
  \includegraphics[scale=1.0]{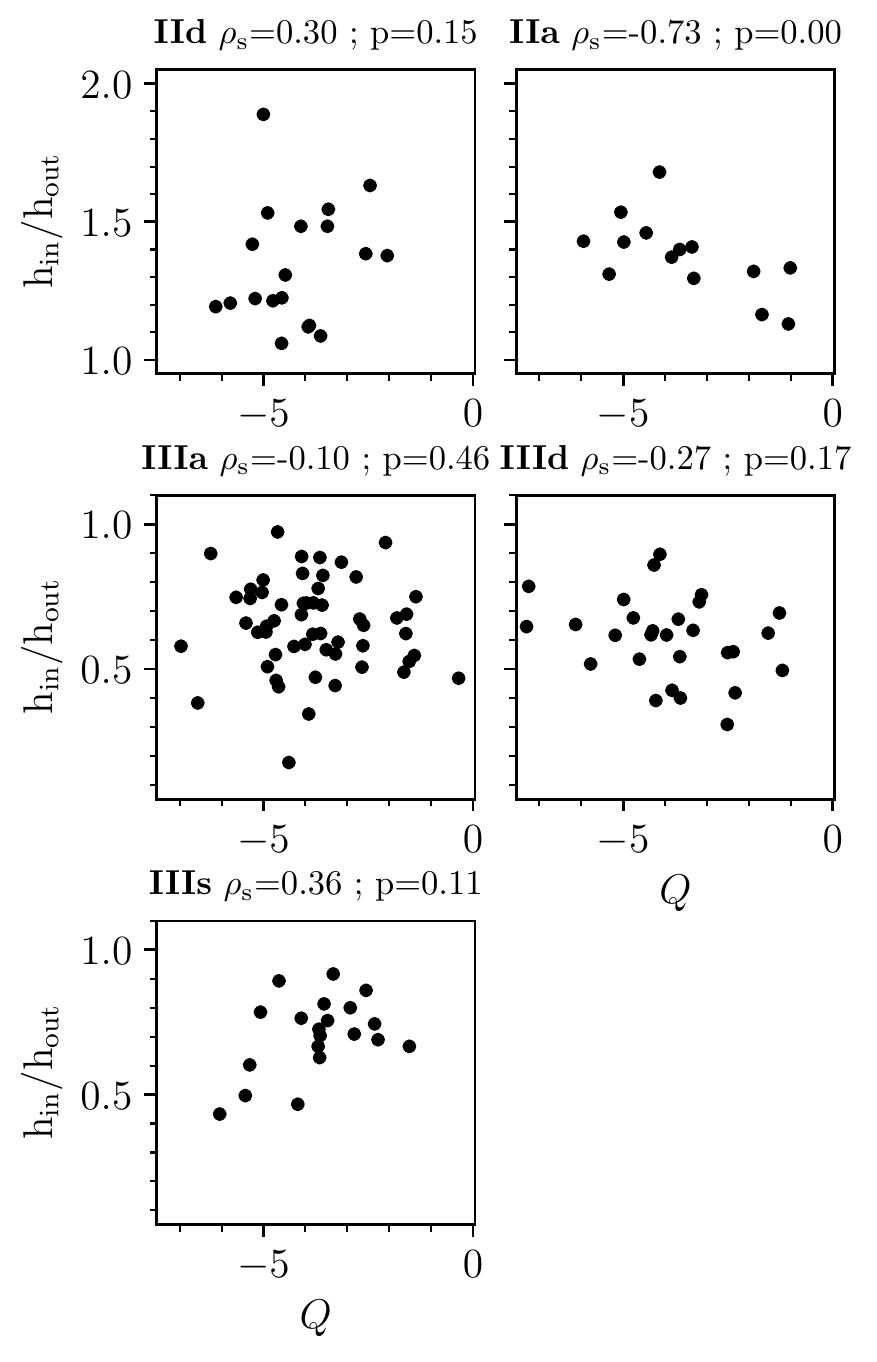}
  \caption{Correlations between break strength and the Dahari
    parameter (see text), separated by outermost break type.  The
    values $\rho_{\rm s}$ denote the Spearman rank correlation
    coefficients for each comparison, while $p$ gives the p-value of
    the Spearman rank test. \label{fig:bstr}}
\end{figure}

As one final consideration regarding the origins of these disk breaks,
we explore the correlation between environment and the outermost break
strength by break subtype.  We define break strength as the ratio of
the disk scalelengths immediately preceding and following the
outermost disk break, h$_{\rm in} /$h$_{\rm out}$ \citep[with scale
  lengths measured using the Numpy routine Polyfit;][]{oliphant06}.
We measure h$_{\rm in}$ between the outermost break radius inward
  to either the second to last break's radius (if present) or to the
  inner radius boundary, and h$_{\rm out}$ between the outermost break
  radius and the outer radius boundary.  Fig.~\ref{fig:bstr} shows the
  correlation between the Dahari $Q$ parameter and the outermost break
  strength in each galaxy in our sample.  For Type~II breaks, higher
values of h$_{\rm in} /$h$_{\rm out}$ denote stronger breaks, while
the inverse is true for Type~III breaks.

As with our previous analysis of environmental parameters, the
correlations we find between environment and break strength are
subtle.  Type~IIa breaks show the strongest correlation, with Spearman
rank correlation coefficient \citep{spearman04} $\rho_{\rm s} =
-0.73$, i.e., stronger Type~IIa disk breaks are found in lower density
environments.  This implies that the tidal features surrounding
galaxies in lower density environments are morphologically sharper
than those in higher density environments.  Studies of the evolution
of intragroup or intracluster light provide a feasible explanation for
this, as tidal streams in higher density environments experience more
frequent torques from passing galaxies and disperse more quickly
  \citep[e.g.,][]{rudick09, janowiecki10}.  Type~IIIa breaks,
however, show no significant correlation with environment, despite
many presumably arising through the same tidal mechanism as Type~IIa
breaks.  In this case, the asymmetric nature of the host's isophotes
may explain this discrepancy: only those tidal features that align
well with the choice of photometry aperture will have a well-measured
scale length (or, indeed, will show a distinct down-bending break at
the feature's peak isophote).

Types IId and IIId show complementary behavior, such that stronger
such breaks are found in higher density environments, however the
trends here are quite weak ($|\rho_{\rm s}| \sim 0.3$) and not
statistically significant (p-values $> 0.15$).  Therefore, while
  the local environment may influence these galaxies' break
  strengths in measureable ways, despite them occupying the lowest
density environments in our sample, the correlations are too weak to
merit further discussion.

Break strengths in IIIs galaxies show a potential positive correlation
with Dahari $Q$, implying stronger breaks in lower density
environments, but again the correlation is small and not statistically
significant ($\rho_{\rm s} \sim 0.36$, p-value $\sim 0.11$).
  This may imply that for IIIs galaxies, the boundary between the
thin disk and the surrounding vertically extended component becomes
weaker in denser environments, but this is again a tentative
  conclusion.

\section{Discussion}

\subsection{Breaks, and what defines the outer disk}

\begin{figure*}
  \centering
  \includegraphics[scale=1.0]{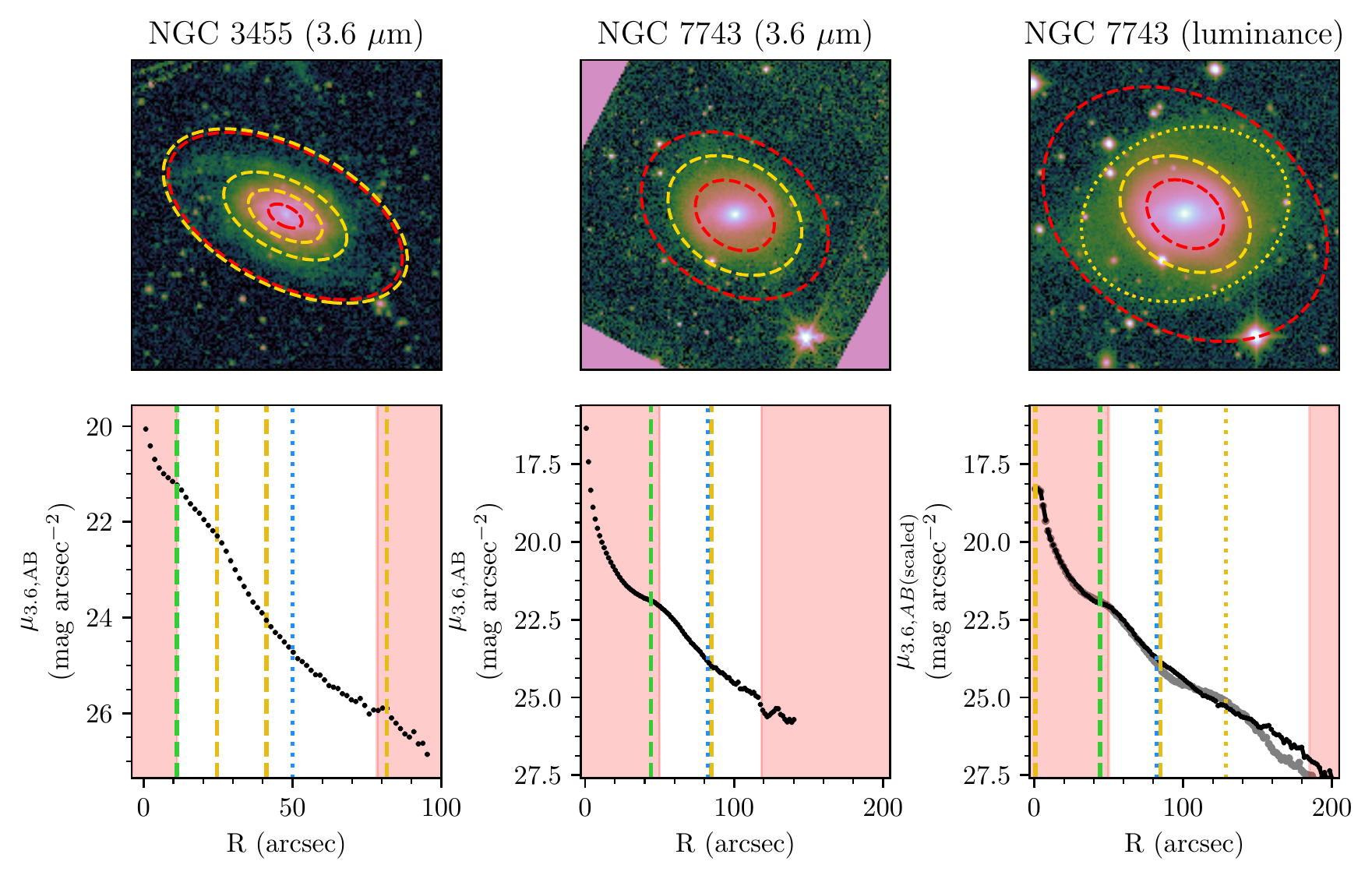}
  \caption{Galaxies ending in Type~IIId disk breaks which seem to show
    faint Type~IId breaks at larger radius.  Ellipses, vertical
      lines, and shaded regions are as in Figures \ref{fig:inrad},
      \ref{fig:btypes1}, and \ref{fig:btypes2}.  {\bf Left:} NGC~3455
    in 3.6$\mu$m, with mild evidence of a Type~IId break occurring
    just beyond the galaxy's faint outer spiral arms.  {\bf Middle:}
    NGC~7743 in 3.6$\mu$m, with no evidence of an outer Type~IId
    break.  {\bf Right:} NGC~7743 in a luminance filter, converted
    roughly to 3.6$\mu$m AB magnitudes by scaling the surface
    brightness profile to the S$^{4}$G profile at the 90\% flux radius
    \citep[image taken as part of the HERON survey;][]{rich17}.  The
    image scale and surface brightness scale are the same as in the
    middle panels.  Black points show the profile using the
      S$^{4}$G isophotal parameters, which are clearly not
      representative of the outer disk; we therefore show a profile
      using a more appropriate position angle and ellipticity in gray
      points as well.  Note that the outer disk contains an extremely
    faint outer ring, at the peak of which is a clear Type~II disk
    break, when the correct isophote shape is used (dotted yellow
      ellipse and vertical line).  The outer radius boundary here
      denotes where the black profile reaches the surface brightness
      of surrounding Galactic cirrus ($\mu_{3.6, \textrm{scaled}} \sim
      27$; this cirrus also causes the additional Type~III break near
      R$\sim$170\arcsec \ in the gray profile). \label{fig:lsbdemo}}
\end{figure*}

By investigating the properties of outer disks using the properties of
the outermost disk breaks of the host galaxies, we have implicitly
defined the outer disk as that which exists beyond the last visible
break radius.  We feel this is a reasonable choice; it is more or less
independent of surface brightness \citep[thereby being defined
  similarly between normal spirals and, e.g., low surface brightness
  spirals;][]{impey97}, and a disk break, by definition, defines a
lasting change in the properties of the disk.  However, even by this
definition, the ``outer disk'' demarcation necessarily depends on the
depth of one's data.  While the S$^{4}$G is the deepest survey of its
kind \citep{sheth10}, disks in galaxies have been seen to extend
beyond the $\mu_{V} = 29$ mag arcsec$^{-2}$ isophote
\citep[e.g.,][]{blandhawthorn05, vlajic11, watkins16}, several
mag arcsec$^{-2}$ fainter than the limits of S$^{4}$G \citep{sheth10,
  salo15}.  It is therefore worth discussing the effects of the
limiting surface brightness on this study, and by extension on all
studies of disk breaks.

We can consider this from the perspective of all of the disk
breaks we have identified.  Because the S$^{4}$G is fairly
  uniform in depth, but includes galaxies with a diversity of central
  surface brightnesses and scale lengths, if Type~III breaks of a
  given subtype typically constitute the outermost breaks of their
  host, we have no reason {\it a priori} to suspect such breaks will
  typically be superseded by additional breaks at larger radius.  By
  contrast, if Type~III breaks of a given subtype were found to be
  frequently superseded by additional breaks within the survey's
  photometric limits, this gives some reason to believe these breaks
  may also be frequently superseded by additional breaks below the
  survey's photometric limits.  Deeper imaging will, of course, be
  necessary to assess the validity of these assumptions, but they
  provide an initial guess.

Of all Type~IIIs and IIIa breaks, 21/24 and 61/89 constitute the
outermost breaks in their hosts, respectively (with 2/21 inner IIIs
breaks followed by additional IIIs breaks, and 6/61 inner IIIa breaks
followed by additional IIIa breaks).  Thus, when these breaks occur,
they tend to persist and so may most often truly demarcate the outer
disks of their host galaxies.  In that regard, it is noteworthy that
of Type~III breaks, Types IIIa and IIIs also appear to occupy the
densest local environments (though again, these trends are fairly
weak).  By contrast, of 96 total Type~IIId breaks identified in
all galaxies in our sample, only 31 persist to the photometric limit
of the data.  In other words, the majority of Type~IIId breaks we
  identify are immediately followed by some other type of break (most
often of Type~IId).  Deeper imaging may therefore reveal that
  few, if any, Type~IIId breaks persist indefinitely.  These Type~III
breaks also showed the weakest correlation with environment, giving
inconsistent results when comparing quite similar measures of the
local surface number density of neighbors; this is expected if
  such breaks do not truly demarcate their hosts' outer disks.

The question also arises whether or not our limit of three breaks per
galaxy may be excluding some breaks beyond the outermost significant
break uncovered by our algorithm.  To this end, we re-ran the
break-finding algorithm on all galaxies with three breaks, splitting
the outermost profiles one additional time to seek out any significant
breaks at larger radius.  We uncovered such breaks in only two cases,
therefore the limit of three breaks per galaxy does not affect our
general conclusions.

If Type~IIIa and IIIs breaks often constitute the outermost
break in their hosts, is this also true of the outermost
Type~IId breaks we find?  Previous studies hint that this may be
the case: for example, in the four disk galaxies with extended
Type~II breaks observed by \citet{watkins14, watkins16}, the surface
brightness profiles beyond these breaks showed no additional changes,
to a limit of $\mu_{B} \sim 30$ mag arcsec$^{-2}$ \citep[but
  see][]{trujillo16}.  Each of these breaks were also coupled to the
reversal of initially blueward color gradients in these
galaxies, suggesting that Type~II breaks followed by redward color
gradients may often mark the outer disk radius in galaxies in which
they are present.  Indeed, the outer disks of such galaxies may be
composed purely of scattered or radially migrated stars
\citep[e.g.,][and many others]{roskar08, schonrich09, minchev11,
  roskar12}, with the break radius marking the extent of spiral
structure and star formation.  Most of the radial color profiles
  of our sample galaxies with extended Type~II breaks did show similar
  initial blueward color gradients up to the break radius; however,
  deeper imaging is necessary to determine if such redward color
  gradients typically follow.  We also found that galaxies with
  outermost breaks of Type~IId appeared to occupy the least dense
environments, a result aligning with that of \citet{erwin12}, who
found (albeit for S0 and Sa galaxies) that Type II breaks tend to
vanish in clusters.  Type~IIa breaks may be similar, although the
persistence of these breaks must depend on the presence or absence of
additional tidal features or distortions present at larger radius.

In terms of representing the true ``outer disk'', the outermost
  Type~IIId breaks in our sample represent the most questionable cases, as
  a significant number of these may show additional breaks
farther out.  We show in Fig.~\ref{fig:lsbdemo} two illustrative
examples: NGC~3455 and NGC~7743.  In NGC~3455, extremely faint outer
disk double spiral arms are visible in the S$^{4}$G image, evidently
the cause of the Type~IIId break.  However, a possible Type~II break
is visible in the surface brightness profile near $\sim$60\arcsec,
just outside of our noise boundary radius, which aligns roughly
with the outermost portions of both arms.  NGC~7743 shows an even
starker example; in the S$^{4}$G image (central panel of
Fig.~\ref{fig:lsbdemo}), the only hint of outer spiral structure comes
from two extremely faint ansae at the northeast and southwest sides of
the visible disk (which occur just beyond the break radius).  However,
deep imaging of the galaxy from the Halos and Environments of Nearby
Galaxies (HERON) survey \citep{rich17} clearly shows two symmetric
spiral arms emerging from the ends of these ansae, at the peak surface
brightness of which is a very clear Type~II break ($\sim 115$\arcsec;
bottom-right panel of Fig.~\ref{fig:lsbdemo}).  This galaxy should
therefore be classified as an oval galaxy of Type~IId, with an outer
disk offset nearly 90$^{\circ}$ from the inner, presumably elliptical
inner disk\footnote{The spiral arms in this galaxy's inner disk wind
  in the opposite direction of those in the outer disk, making this an
  extremely peculiar galaxy.}.  None of this structure is visible in
the S$^{4}$G imaging, and only a hint of it is visible in SDSS
imaging.

This is worth considering in any future disk break surveys.
Particularly for surveys like SDSS, which do not probe to particularly
deep surface brightness limits, Type~IIId galaxies may prove a
significant contaminant if one simply classifies breaks as Types I,
II, and III, with no further consideration of these breaks' possible
origins.  With these caveats in mind, we consider now the origins of
Type~III breaks in general.

\subsection{The varying origins of Type~III disk breaks}

Of all galaxies with Type~III breaks that persist to the photometric
limit of our data, $\sim$53\% are related to distorted isophotes or
tidal features.  In these cases, a radial profile assuming azimuthal
symmetry is simply not an accurate representation of the true behavior
of the galaxy's isophotes.  These disk breaks are therefore as much
methodological in origin as they are physical in origin.  This is
reflected as well in the lack of correlation between Type~IIIa break
strength and the Dahari $Q$ parameter (Figure \ref{fig:bstr}).

Given that such breaks are partly methodological in origin, one
  may question whether they should be considered as ``disk breaks'' at
  all \citep[e.g.,][in the case of their Type~II-AB
    classification]{pohlen06}.  We include them here for the sake of
  comparison with \citet{laine14} and L16, but it is clear that the
  scale lengths, break radii, and other physical parameters derived
  for these features depend strongly on the choice of aperture used to
  measure them.  Particularly when seeking out correlations involving
  break strength, break radius, or similar parameters, such galaxies
  should always be classified separately.  Regardless of their status
  as break-hosting galaxies, however, the source of their asymmetry is
  worth considering.

While in many cases the asymmetric nature of the Type~IIIa break hosts
is almost certainly the remnant of a tidal interaction or merger event
(e.g., NGC~474, NGC~2782, NGC~4651, etc.), lopsidedness in galaxies
need not always emerge through environmental influence
\citep[e.g.,][]{zaritsky13}.  That said, we have found that such
distorted galaxies (of Types IIIa and IIa) may occupy denser local
environments than all break type hosts but those of Type~IIIs in this
galaxy sample.  We also found, for Type~IIa galaxies, a strong
anti-correlation between break strength and the tidal Dahari
  parameter $Q$.  This is intuitive if such features are indeed most
frequently tidal in origin: extended tidal streams preferentially form
through low-speed, prograde encounters \citep[e.g.,][]{toomre72,
  eneev73, farouki82, barnes92}, such as might be more common in
low-mass groups.  Tidal features are also more easily disrupted and
dispersed in denser environments \citep{rudick09, janowiecki10},
transforming previously sharp or shell-like tidal streams into more
extended and diffuse streams.  Therefore, we would expect to more
often find these types of tidal signatures in galaxy pairs and groups
than in either isolated or cluster galaxies.

Our firmest conclusion is that galaxies with outermost breaks of
Type~IIIs occupy the highest density environments.  If these breaks
mark the transition to vertically hotter galactic components, this is
also intuitive.  Galaxies occupying higher mass group or cluster halos
should instead experience more frequent, higher velocity encounters
\citep[galaxy harassment;][]{moore96} and ram-pressure stripping
\citep{spitzer51, gunn72}, mechanisms capable of contributing to the
morphology-density relationship \citep{spitzer51, oemler74,
  dressler80, cappellari11} by transforming spiral galaxies into
lenticulars or ellipticals \citep[e.g.,][]{moore96, boselli06,
  cappellari11}.  Strong and sustained such harassment would heat the
host's thin disk, thereby both expanding this surrounding spheroidal
component and weakening the apparent boundary between this component
and whatever remains of the thin disk.  We see some evidence of this
as well, in a possible correlation between break strength and
$Q$ for Type~IIIs galaxies (Fig.~\ref{fig:bstr}).  The
spheroidal components of many Type~IIIs disk galaxies may thus be
merely stellar populations tidally heated by such harassment, and
therefore may represent examples of galaxies undergoing the transition
from late to early type.

We may also see some evidence of this process in these galaxies' color
profiles, which are split between those with blue inner disks and
redward color gradients, and those with red ($g-r \sim 0.7$) inner
disks and flat, only slightly bluer ($g-r \sim 0.6$) outer disks.  This
behavior is reflected in these galaxies' morphological types as well,
which is bimodal: $11/21$ galaxies have T-types $\leq 2$, with the
remainder $\geq$4, six of which are $\geq$8 \citep{buta15}.  By
contrast, the other break type hosts show roughly Gaussian T-type
distributions peaking around $\sim$3 (though Type~IIIa disks also show
a sharp peak above 8), hence early type disks are relatively more
prevalent among Type~IIIs galaxies in this sample.

The gas in outer disks is vulnerable to both tidal forces \citep[which
  may torque the gas, leading to infall; e.g.,][]{barnes96, hopkins13}
and ram-pressure stripping \citep{gunn72}, both of which can deplete
gas and truncate star formation in the outer regions of galaxies
occupying dense environments.  Gas that falls inward due to tidal
torquing can go on to ignite central starbursts
\citep[e.g.,][]{barnes91, hernquist92, barnes96, cox06, hopkins13},
which may be the origin of those IIIs galaxies with blue central
regions.  If the gas is not replenished, sustained such interactions
will deplete the galaxy's reservoir and halt star formation.  Rapid
episodes of central star formation can increase both the central
metallicity and dust content, leading to quite red integrated colors
once star formation ceases.  The two populations of Type~IIIs galaxies
in this sample may thus reflect galaxies currently undergoing
quenching, and galaxies that have already quenched.

Simulations and observations suggest short quenching
timescales, however \citep[50 Myr -- 2 Gyr for massive galaxies,
  e.g.,][]{quai18, wright18}.  At the upper end of these estimates,
assuming typical ages of 10 Gyr, one might expect $\sim 20$\% of any
random population of galaxies undergoing quenching to still show
significant star formation at $z=0$.  In our sample, $6/16
\sim$40\% have blue inner disks, quite high even for this upper limit
estimate.  This suggests that other factors may be at play as well in
the formation of Type~IIIs disk breaks.  The galaxies NGC~4800 and
NGC~5676 provide interesting such examples: both galaxies appear as
quite normal-looking star-forming disks embedded within a more
diffuse, extended stellar envelope, and both galaxies show blueward
color gradients that reverse at the IIIs break radius.  These galaxies
may instead represent galaxies with extended thick disks, following
the suggestion by \citet{comeron12}.  If these thick disks arose
during the early stages of these galaxies' evolution \citep[as is
  thought to be true of the Milky Way; e.g.,][]{bensby05}, they need
not be undergoing quenching at all.  If this is the case, we would
expect that with deeper imaging, the fraction of IIIs galaxies would
increase, yet the correlation between IIIs breaks and environment may
weaken, as only those Type~IIIs breaks that occur at high surface
brightness may be related to galaxy harassment and tidal interactions.

Previous work using the surface brightness profiles of S0
  galaxies has also failed to find any difference in the Type~III disk
  fraction between field and cluster environments
  \citep[e.g.,][]{erwin12, pranger17, silchenko18}.  Several
  possibilities may explain this, in light of our conclusions here
  regarding Type~IIIs galaxies.  First, no distinction was made in
  these studies between Type~III break subtypes, and Type~IIIs breaks
  are relatively rare \citep[$\sim$30\% of all Type~III breaks found
    in][]{erwin12}.  Second, these studies made no distinction between
  the field and galaxy groups, only between galaxies inside and
  outside of clusters; the correlations between break subtypes and
  environment we find in our study are local.  Third, many of the Type
  IIIs galaxies in our sample may not be transitioning into S0
  galaxies, but into rotation-supported ellipticals.  Differentiating
  among these scenarios to understand the origins of these Type~IIIs
  galaxies will therefore require a much more detailed study,
  including detailed examination of stellar populations and
  kinematics.

The origins of Type~IIIa and IIIs breaks thus appear diverse, but lean
toward environmental influence.  The remaining Type~III galaxies in
this sample are of Type~IIId, suggesting their outer isophotes are
dominated by extended, symmetric spiral arms or star-forming
disks.  Their color profiles appear to uphold this interpretation:
most show roughly continual blueward color gradients (with a
sharper downturn at the break radius; Fig.~\ref{fig:cprofs}).  This
suggests a link between these galaxies and those of Type~IId, which
also often show blueward color profiles up to their own
downbending break radii.  However, the comparison between these
populations' local environmental density was ambiguous: in only one of
three density parameters ($\Sigma_{3}^{A}$) did the two local
environments appear similar.

While surface brightness limitations seem to play a role (as discussed
in the previous section), some of this ambiguity may also be inherent
to the classification of Type~IIId galaxies itself.  While Types IIIa
and IIIs have quantifiable signatures (the $m=1$ Fourier mode
amplitude and the isophotal ellipticity gradient, respectively),
classifying Type~IIId galaxies relies mostly on visual inspection and
comparison across wavelengths.  We also took a conservative approach
to classification, labeling as part of the ``disk'' any largely
symmetric spiral features, including those with large pitch angles
that may be tidal in nature (e.g., NGC~2854, shown in
Fig.~\ref{fig:btypes1}).

Whether or not extremely faint, extended spiral arms like those of
NGC~7743 (Fig.~\ref{fig:lsbdemo}) are tidal in nature is also
unclear.  NGC~7743 itself does lie at roughly the same redshift
distance as NGC~7742 \citep{devau91}, and is separated by only
$\sim$50\arcmin \ in projection ($\sim$320 kpc at 20 Mpc).  NGC~7742
contains no bar, yet still hosts a star-forming nuclear ring
\citep{comeron10}, as well as a counter-rotating gas disk
\citep{silchenko06}, implying a past merger event.  These galaxies'
local environment thus is or recently has been somewhat active,
implying the outer disk of NGC~7743 may also be tidal in origin.  Two
other galaxies in our sample with similar faint outer rings also have
nearby companions: NGC~3684, near NGC~3607, and NGC~4151, near
NGC~4145 \citep{devau91}.  However, M94 (not in this sample), host to
another oval and faint outer ring, is one of the most isolated
galaxies in the nearby universe \citep{smercina18}.  What we have
labeled as ``disklike'' outer isophotes may therefore contain a mix of
both secular features and tidal features, which are not easily
distinguishable.  Indeed, some breaks of this type may also
  result from a combination of secular processes and environmental
  interaction \citep[e.g.,][]{clarke17, ruizlara17}.  Again, this
implies that the inclusion of Type~IIId galaxies with other Type~III
galaxies will serve mainly to muddle any broad conclusions one may
come to regarding the origins of Type~III disks.

\subsection{Implications for future work}

The trends we have uncovered here, though weak, are promising.  We
have found multiple lines of evidence that the majority of 
  upbending breaks in disk galaxy surface brightness profiles are
indeed created through environmental influences, either through the
creation of extended tidal features that skew the surface brightness
profile at large radius (which may not be best characterized as
  ``disk breaks''), or through merging or harassment in dense
environments creating vertically hot, non-star-forming stellar
components in the host disks.

Our current study, however, suffers from quite small sample sizes.
This combined with any inherent subjectivity bias in our break
categorization scheme --- for example, our difficulty in
  differentiating symmetric tidal features from loosely wound outer
  spiral arms, as in the case of NGC~7743 described above --- makes
these conclusions still quite tentative.  However, it is worth asking
if such a detailed analysis is necessary to disentangle the competing
influences on break formation in order to come to robust conclusions
across studies regarding, e.g., the influence of secular
vs. environmental processes on break formation.  To test this, we
applied the BEST analysis and the Mann-Whitney U test on the full
dataset from L16, split generally into Type~I, II, and III galaxies
using these authors' break classifications.  We then performed
  pairwise comparisons between populations classified by break type
  (e.g., Type~I vs. II, II vs. III, and I vs. III).  In all
comparisons save one, we found significant differences between
populations' mean Dahari $Q$ parameter or $\Sigma_{3}^{A}$ (at the
$\gtrsim$95\% confidence level), with Type~III break hosts occupying
the highest density environments.  Only a comparison between Type~III
and II break hosts using the Dahari $Q$ parameter failed to uncover
any significant difference using these broad categories.  However, if
we eschew all breaks from the Type~II sample labeled II.i
\citep[denoting breaks inside the bar radius;][]{laine14}, we do find
significant differences between the populations (CP $\sim$95\%), with
again Type~III galaxies occupying the higher density
  environments.  This suggests that the presence of disk breaks
inside the bar radius are influenced far more by the presence of the
bar than by any environmental factors, hence the inclusion of such
breaks in a sample of Type~II disk break hosts will again muddle any
conclusions one may come to regarding the role of environment on
disk break formation.

Some level of detail does therefore appear important regarding disk
break classifications.  In future studies of disk breaks, we therefore
recommend specificity beyond simply identifying whether or not each
galaxy hosts a Type~II or Type~III disk break at some radius. 
  Depending on what physical scenarios are being explored using the
  disk breaks, perhaps this need not be as complex as what we have
done in this study.  Still, clarity should come through at least
token effort at identifying the causal features behind the disk breaks
one identifies, so that, to the best of one's ability,
environmental effects may be paired with environmental effects,
and secular effects may be paired with secular effects.  Also,
  one should be mindful that the ``breaks'' one is measuring are
  actually breaks and not simply a reflection of distorted isophotes.
Doing otherwise risks cloudying already subtle correlations and
scaling relations that might be illuminated by disk breaks.

\section{Summary}

We have re-analyzed the sample of Type~III break-hosting galaxies
first identified by \citet{laine14, laine16} in order to better
understand the physical origins of Type~III disk breaks.

Using an unbiased break-finding algorithm, we have
reclassified the disk breaks in each of these galaxies and tied each
significant change in slope to distinct features across the host
disks.  We find that, if one considers all breaks in all sample
galaxies, Type~III breaks are most frequently related to either
morphological asymmetry (Type~IIIa) or to features such as
extended spiral arms or star forming regions (Type~IIId).  Many
of the latter breaks, however, do not persist to the photometric
limit of the data, typically being superseded by either
asymmetry-related breaks, breaks associated with an outer spheroidal
component (Type~IIIs), or mild down-bending (Type~II) breaks
associated with, e.g., the ends of spiral arms (Type~IId).  Type
  III breaks associated with extended spiral structure or star
  formation (Type~IIId), therefore, may not often truly characterize
  their hosts' outermost disks.

We have analyzed the stellar populations of these galaxies using their
$g-r$ color profiles, separating galaxies by outermost break type (the
outermost break being the most sensitive to the local
environment).  We find that those galaxies with Type~II breaks in
their outskirts show blueward colors gradients up to the disk
  break radius; deeper multiband imaging is needed to determine
  whether or not these disks typically show redward color gradients
  beyond the break radius, as in Type~II break galaxies from the
  studies by \citet[e.g.,][]{bakos08, zheng15}.  Galaxies with
asymmetry-related breaks (Type~IIIa) show flat and somewhat
  blue outskirts, possibly related to enhanced outer disk star
formation caused by the asymmetry.  Galaxies with outer spheroids
  (Type~IIIs) show two distinct classes of color profiles: those with
blue central regions and red outskirts, and those with red central
regions and flat, only slighter bluer outskirts.  This suggests that
whatever mechanism has created the spheroids in these galaxies may
also be related to quenching, with torqued gas falling inward,
enhancing central star formation and enriching the core ISM in metals
and dust.  However, because the fraction of such galaxies with blue
cores is too high compared to simulated populations of quenching
galaxies, some of these outer spheroids may also be thick disks with
extended scale lengths surrounding otherwise normal spirals.

Additionally, we compared the local environments of the various
break-type hosts using three different parameters: the Dahari $Q$
parameter (a measure of tidal strength), and two measures of the local
surface density \citep[within the third nearest-neighbor projected
  radius, and within the groups defined by][]{kourkchi17}.  Though
sample sizes are small, and therefore conclusions are still tentative,
we find several consistent trends.  We find that galaxies with outer
spheroids (Type~IIIs) seem to occupy the highest-density
environments in our sample, suggesting again a connection between
these components and, e.g., galaxy harrassment and quenching
mechanisms.  Along this line, we find a possible positive correlation
between these galaxies' break strengths and Dahari $Q$
parameter, suggesting that the boundary between the thin disk and the
outer spheroid in such galaxies becomes weaker in denser environments.
Galaxies with asymmetries (Types IIIa and IIa) appear to occupy
the next highest-density environments, suggesting that most of these
asymmetries are indeed related to tidal disturbances, albeit not as
violent of tidal disturbances as those leading to the outer spheroids.
Galaxies with down-bending breaks in their outskirts which seem
associated with spiral arms or star-forming regions (Type~IId)
appear to inhabit the least-dense environments, suggesting that such
breaks only persist in mostly undisturbed galaxies, a result in line
with the near absence of Type~II disks in clusters \citep{erwin12}.
Finally, we find little concrete correlation between environment and
those Type~III breaks associated with spiral arms and extended star
formation (Type~IIId).  We have shown that some of this
ambiguity arises from insufficient surface brightness depth, as in
several cases (most intriguingly, NGC~7743) we find that these
galaxies' outer disks show Type~II breaks at lower surface brightness.

In general, most of those Type~III disk breaks that occur in the outer
disk appear to have arisen through environmental influences rather
than secular evolution.  Therefore, some distinction should be made in
future studies regarding the physical origins of these breaks,
specifically break location (inner disk vs. outer disk) and any
associated morphological features (e.g., bars, spiral arms, tidal
streams, or outer spheroids).

\begin{acknowledgements}
  Financial support for this project was provided by the Academy of
  Finland (grant No. 257738).  We thank R. Michael Rich for the use of
  his HERON imaging.  This paper has benefited from discussions within
  the SUNDIAL ITN network, an EU Horizon 2020 research and innovation
  program under the Marie Sk\l{}odowska-Curie grant agreement
  No. 721463. J.J. and H.S. also acknowledge support from the SUNDIAL
  ITN network.  This project made use of the following astronomical
  and mathematical software packages: AstroPy \citep{astropy:2013,
    astropy:2018}, MatPlotLib \citep{hunter07}, PyMC3
  \citep{salvatier16}, Pandas \citep{mckinney10}, NumPy
  \citep{oliphant06}, and SciPy \citep{jones01}.  We wish to thank as
  well the S$^{4}$G team for providing the imaging data and data
  products used in this research.  We acknowledge the usage of the
  HyperLeda database (http://leda.univ-lyon1.fr).  Finally, we
    wish to thank the referee, Michael Pohlen, for his careful and
    thorough review of our paper.
\end{acknowledgements}

\bibliographystyle{aa}
\bibliography{refs}

\begin{appendix}

\section{The influence of systematics on surface brightness profiles}

\begin{figure*}
  \centering
  \includegraphics[scale=1.0]{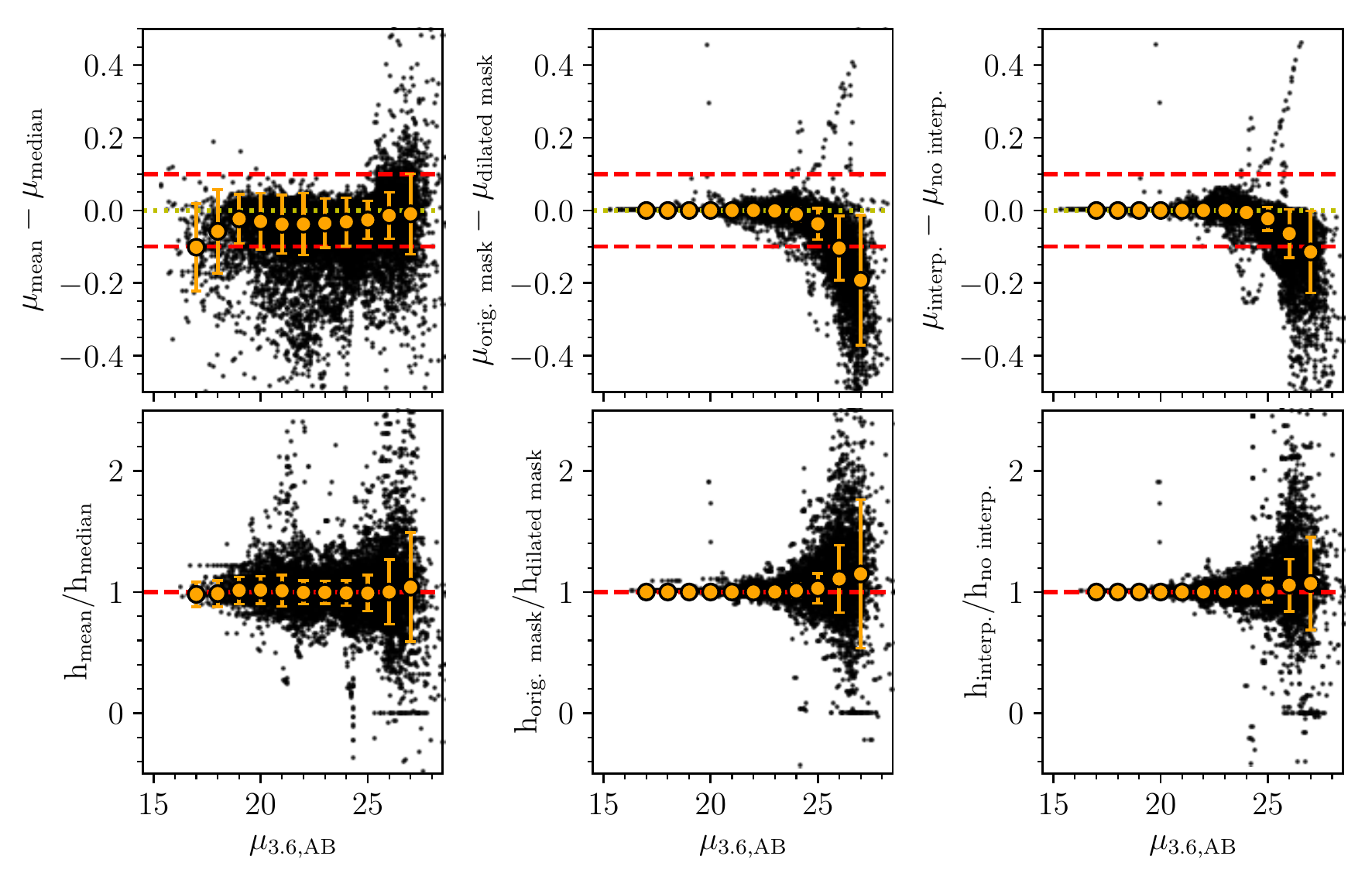}
  \caption{Showcasing systematic effects of different methods on the
    derivation of surface brightness as a function of radius, and
    local slope as a function of radius.  Top row panels are, from
    left to right (all plotted as a function of surface brightness):
    difference between mean and median surface brightnesses;
    difference between surface brightnesses measured using the
    original S$^{4}$G masks from \citet{munozmateos15} and those
    measured using dilated versions of the same masks; and surface
    brightnesses measured through interpolation across masks vs. those
    measured ignoring masked pixels.  Bottom row panels show the same
    comparisons of systematics, but for the ratio of the measured
    local slopes (converted to scale length) using the varying
    methods.  Orange points show the running median, with errorbars
    showing the interquartile range at each surface brightness bin.
    In the top row panels, red dashed lines show $\pm$0.1 mag
    arcsec$^{-2}$.  In the bottom row panels, red dashed lines show a
    slope ratio of 1 (no difference between
    methods).  \label{fig:systematics}}
\end{figure*}

We explore here the effect of three different systematics on our
measured surface brightness profiles: the choice of mean
  vs. median as the average flux in the radial bin, masking, and
interpolation across masks vs. ignoring masked pixels.
Fig.~\ref{fig:systematics} shows a summary of the trends.  In all
panels, we plot the difference between methods as a function of
surface brightness in black points, showing all measured surface
brightnesses or scale lengths at all radial bins in all 175 sample
galaxies.

\subsection{Mean vs. median}

In the leftmost panels of Fig.~\ref{fig:systematics}, we show the
effects of using the mean vs. median surface brightness.  We show the
effect on surface brightness in the top-left panel.  In general, the
mean surface brightness is brighter than the median, such that
$\mu_{\rm mean} - \mu_{\rm median} < 0$.  Regardless of surface
brightness, this difference is $\sim -0.03$ magnitudes, with long
tails to more extreme negative values (as large as half a magnitude in
some cases).  Below surface brightnesses of $\mu_{3.6, \textrm{AB}}
\sim 25.5$, the scatter becomes symmetric and the median difference
shifts closer to 0 (though is still slightly negative).  This is
roughly the limiting surface brightness of the S$^{4}$G data, based on
the uncertainty in the sky subtraction \citep{salo15}.  At lower
surface brightness, both mean and median begin to trace the peak of
the background sky's flux distribution (with a $\sim$0.01 mag
arcsec$^{-2}$ skew to higher fluxes when using the mean), with large
scatter.  Where the galaxy's light dominates the isophotes, then, the
mean will tend to more closely trace the brightest features in the
disk.  In the sky-dominated regime, the mean will skew toward the
brightest unmasked features \citep[which, in the case of unresolved
  imaging, will usually be unmasked background galaxies or foreground
  stars; e.g.,][]{rudick10}, and therefore is the more biased
estimator at low surface brightness unless extreme care is taken to
remove all sources of foreground or background contamination.

We show the effect of mean vs. median on the local slope in the
bottom-left panel.  We find no systematic effect on slope regarding
the mean vs. median surface brightness, though again the scatter is
large at all surface brightnesses.  However, most of the outliers in
this case skew to high values, implying that mean surface brightnesses,
in extreme cases, tend to yield larger scale lengths.  Unlike the
effect on surface brightness, the effect on slope does not become
symmetric below the noise limit.  Likely this is due to the
flux-weighted nature of the mean, which in these outlier cases tends
more often to skew toward the brightest peaks in surface brightness,
thereby leveling out the profile.

A detailed comparison of the mean vs. median profiles can be found on
the website accompanying this paper
(\url{https://www.oulu.fi/astronomy/
 S4G_TYPE3_DISC_BREAKS/breaks.html}).  Median surface brightness
profiles are included there alongside the mean profiles for each
galaxy as gray dashed lines.  We also provide these profiles in
tabular form via CDS.

\subsection{Masking}

In the two center panels of Fig.~\ref{fig:systematics}, we display the
effects of masking.  In the top panel, we show the difference between
surface brightnesses measured using the original S$^{4}$G masks from
\citet{munozmateos15} and those measured using the same masks but
grown through dilation with a 3-pixel radius kernel.  Down to roughly
$\mu_{3.6, \textrm{AB}} \sim 24$ mag arcsec$^{-2}$, the measured
surface brightnesses appear more or less insensitive to masking, on
average (with IQR$<$0.01 mag arcsec$^{-2}$), though in
individual cases the induced change can be as high as 0.1 mag
arcsec$^{-2}$.  At lower surface brightnesses, the masking begins to
affect the profile systematically, with harsher masking resulting in
fainter measured surface brightnesses and increased profile-to-profile
scatter.  By $\mu_{3.6, \textrm{AB}} \sim 25$, the median profiles
have altered by $>$0.1 mag arcsec$^{-2}$, with a scatter of roughly
the same value.  Visual inspection of the images reveals that most of
this effect can be attributed to the wings of the PSFs of bright stars
leaking through masks, which artificially elevate the mean surface
brightness measured in annuli where the galaxy isophotes are faint.
We also find that the median, while more robust to masking, is not
immune to this effect.

We show the effect of masking on the slope in the center-bottom panel.
As discussed in Section 2, masking affects the slope profiles most at
low surface brightness by increasing the measure scale lengths.  The
effect becomes strongest near $\mu_{3.6, \textrm{AB}} \sim 26$ mag
arcsec$^{-2}$, though scatter is already significant at $\mu_{3.6,
  \textrm{AB}} \sim 25$.  The overall influence of insufficient
masking was to induce several spurious Type~III breaks in our
profiles, therefore we chose to include only those breaks that were
not altered by the choice of mask.  Again, use of the median over the
mean alleviates but does not remove this effect.  Proper masking is
critical to measuring accurate surface brightness profiles at low
surface brightnesses.

\subsection{Interpolation across masks}

Finally, in the rightmost panels of Fig.~\ref{fig:systematics}, we
show the effect of our choice to interpolate across masks for this
study.  Interpolation was done primarily so that the surface
brightness profiles could be used to measure the mass and potential of
each galaxy as a function of radius \citep[for example, for the
  purposes of measuring the influence of the bar;
  e.g.,][]{diazgarcia16a, diazgarcia16}.  For the sake of comparison
with \citet{laine14, laine16}, we used the same surface brightness profiles,
which included this interpolation.  This, too, however, carries a
systematic effect, such that interpolation artificially increases the
surface brightness in a given radial bin.  This effect is linked to
the effect of masking, however; light leaking through masks carries
through the interpolation, enhancing the surface brightness in the
masked region.  Use of more aggressive masks also removes this
effect.

We show the effect of interpolation on the measured slopes in the
bottom-right panel.  Again, there is a slight systematic in the same
direction as that induced by the masking.  The root cause is again the
same, therefore use of appropriately aggressive masks removes the
systematic effects of interpolation as well.

In general, we find that these systematic effects alter our
conclusions only in the low surface brightness regime, with masking
and interpolation across masks showing the greatest influence.  As
discussed in Section 2, we have used only the larger masks in our
analysis, thereby alleviating both of these biases.  While individual
galaxies can show extreme changes under these various systematics,
overall the effects are quite small and so do not alter any of the
conclusions we have discussed regarding populations of galaxies.

\end{appendix}

\end{document}